\pdfoutput=1
\documentclass[cernpreprint,atlasdraft=false,texlive=2019,UKenglish,texmf]{atlasdoc}
 
\usepackage{atlaspackage}
\usepackage{atlasbiblatex}
 
\usepackage[process,jetetmiss]{atlasphysics}
 
\usepackage{multirow, makecell}
 
\graphicspath{{logos/}{figures/}}
 
\usepackage{ANA-EXOT-2018-33-PAPER-defs}

\AtlasTitle{Search for type-III seesaw heavy leptons in dilepton final states
in \(pp\) collisions at \( \sqrt{s} = 13\,\text{TeV} \) with the ATLAS detector}
 
\hypersetup{pdftitle={Search for type-III seesaw heavy leptons}, pdfauthor={The ATLAS Collaboration}}
 
\AtlasAbstract{
A search for the pair production of heavy leptons as predicted by
the type-III seesaw mechanism is presented. The search uses proton--proton  
collision data at a centre-of-mass energy of 13\,TeV, corresponding
to \( 139\,\text{fb}^{-1} \) of integrated luminosity recorded by the ATLAS
detector during Run 2 of the Large Hadron Collider.\@
The analysis focuses on the final state with two light leptons (electrons
or muons) of different flavour and charge combinations, with at least
two jets and large missing transverse momentum. No significant excess over
the Standard Model expectation is observed.
The results are translated into exclusion limits on heavy-lepton masses,
and the observed lower limit on the mass of the type-III seesaw heavy leptons
is 790\,GeV at 95\,\% confidence level.  
}
 
\author{The ATLAS Collaboration}
 
\AtlasRefCode{EXOT-2018-33}

\AtlasJournal{\EPJC}
\AtlasJournalRef{Eur. Phys. J. C 81 (2021) 218}
\AtlasDOI{10.1140/epjc/s10052-021-08929-9}
 
\PreprintIdNumber{CERN-EP-2020-130}
 
\begin{document}
 
\maketitle
 
\tableofcontents
 
\section{Introduction}
\label{sec:intro}
 
The experimental observation of neutrino oscillation provides conclusive evidence
that neutrinos have non-zero masses which are much smaller than those of the
charged leptons~\cite{Tanabashi:2018oca}.
While in the Standard Model (SM) the charged fermions acquire masses by coupling to the
Higgs boson, the origin of neutrino masses is still unknown.
An elegant extension of the SM, accounting for the smallness of neutrino masses compared
to other fundamental particles, is the seesaw mechanism~\cite{Minkowski:1977sc,yanagida1979proc,glashow1980proceedings,gell1980supergravity,Mohapatra:1979ia}.
It introduces a neutrino mass matrix with Majorana mass terms and very small neutrino mass eigenvalues, emerging from
the existence of a heavy neutrino partner for each of the light neutrino species.
Several models implementing the seesaw mechanism exist
(see e.g.\ \Refn{\cite{Ma:1998dn}} for a concise overview) with different
particle content. Among these, the type-III model~\cite{Foot:1988aq} introduces
at least one extra triplet of heavy fermionic fields with zero hypercharge
in the adjoint representation of SU(2)\(_\text{L}\) which couple to 
electroweak (EW) gauge bosons and generate neutrino masses through Yukawa
couplings to the Higgs boson and neutrinos. Consequently, these new charged and
neutral heavy leptons could be produced via EW processes in proton--proton (\(pp\))  
collisions at the Large Hadron Collider (LHC).
 
Several \typeIIIseesaw  heavy-lepton searches have been performed in different decay channels.
The search presented in this paper is an extension of a similar search performed by ATLAS in \RunOne at \(\rts = \SI{8}{\TeV}\)~\cite{EXOT-2014-07}, using the same final state of two light leptons and two jets, which excluded heavy leptons with masses below \SI{335}{\GeV}. In \RunOne  this search was complemented by another ATLAS search for heavy leptons using the three-lepton final state~\cite{EXOT-2014-08}, which excluded heavy-lepton masses below \SI{470}{\GeV}.
A search by the CMS Collaboration using their full \RunTwo dataset at \(\rts = \SI{13}{\TeV}\)~\cite{CMS-EXO-19-002} was performed, focusing on final states with at least three leptons, setting limits on the mass of \typeIIIseesaw heavy leptons of up to \SI{880}{\GeV}. The analysis presented in this paper, based on the full ATLAS \RunTwo dataset, is a novel measurement performed by an LHC experiment in \RunTwo searching for \typeIIIseesaw heavy leptons using the dilepton signature.
Substantial refinements relative to the published ATLAS \RunOne analysis have been made in the background estimation and selection of signal candidate events, as well as in the theoretical signal modelling and signal cross-section next-to-leading-order calculation, all of which have a significant impact on the achieved measurement sensitivity of this analysis.
 
A minimal \typeIIIseesaw model
is used to optimise the analysis strategy and interpret the search results,
as in \Refn{\cite{Biggio:2011ja}}.
This model introduces a single fermionic triplet with unknown (heavy) masses
for one neutral and two oppositely charged leptons, denoted by \((\Lp, \Lm, \Nz)\).
Here \Lp is the antiparticle of \Lm and \Nz is a Majorana particle.
These heavy leptons decay into a SM lepton and a \Wboson, \Zboson or \Hboson boson.
The heavy leptons are assumed to be degenerate in mass. This assumption does
not limit the generality of results within the context of \typeIIIseesaw models because the theoretical calculations predict only a small
mass-splitting due to radiative corrections and the resulting transitions between
heavy leptons are highly suppressed~\cite{Franceschini:2008pz}.
 
The remaining degrees of freedom in the simplified model are the unknown mixing
angles \( V_\alpha \) (\( \alpha = e,\mu,\tau \)) between the SM leptons and
the new heavy-lepton states, which enter only in the expressions for the \Lpm
and \Nz decay widths. The production cross-section does not depend on
the \( V_\alpha \) parameters because the heavy leptons are produced through
the coupling to the EW bosons. Only the branching fraction \(\mathcal{B}_\alpha \) of
the \Lpm and \Nz decays to lepton flavour \( \alpha \) depends on the values
of the mixing angles and is proportional to
\(\mathcal{B}_\alpha = |V_\alpha|^2/(|V_e|^2 + |V_\mu|^2 + |V_\tau|^2) \).
In this analysis the decay branching ratios are assumed to be equal for all three
lepton flavours, with \(\mathcal{B}_e = \mathcal{B}_\mu = \mathcal{B}_\tau = 1/3 \).
 
\begin{figure}[tbp]
\centering
\subfloat[]{
\includegraphics[width=0.4\textwidth]{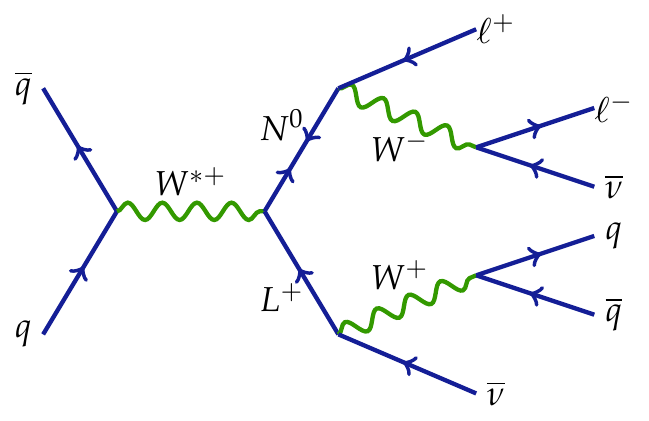}
\label{fig:feynman-OS}
}
\subfloat[]{
\includegraphics[width=0.4\textwidth]{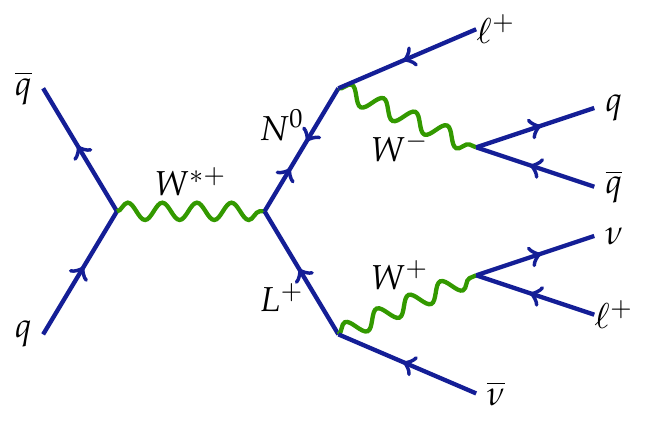}
\label{fig:feynman-SS}
}
 
\caption{Feynman diagrams of the dominant production process in the
\protect\subref{fig:feynman-OS} opposite-charge and
\protect\subref{fig:feynman-SS} same-charge final-state cases.
}
\label{fig:feynman}
\end{figure}

The dominant production mechanism for \typeIIIseesaw heavy leptons in \pp
collisions is pair production through \( q\bar{q} \rightarrow W^{*} \rightarrow \Nz \Lpm \),
and the largest branching fraction is the one with two \Wboson bosons in the
final state: \( \Nz \rightarrow \Wpm \ell^\mp \, (\ell=e,\mu,\tau) \), and
\( \Lpm \rightarrow \Wpm \nu \, (\nu =\nu_e,\nu_\mu,\nu_\tau) \).
Branching fractions through intermediate \Wboson bosons range from
approximately \SI{80}{\%} down to approximately \SI{60}{\%} over the
heavy-lepton mass range. This search is optimised for the dominant
processes \( \pp \rightarrow \Nz \Lpm \rightarrow \Wpm \ell^\mp \Wpm \nu \),
where one \Wboson boson decays leptonically and the other decays hadronically
(\Fig{\ref{fig:feynman}}). However, all \( \pp \rightarrow \Nz \Lpm \) and
\( \pp \rightarrow \Lpm \Lmp \) processes with two or more leptons in the final
state are considered to form part of the signal. Only final states containing
electrons and muons are considered, including those from leptonic \( \tau \) decays.
 
The exclusive topology of the final state consists of two jets from
the hadronically decaying \Wboson boson, large missing transverse momentum
and a lepton pair with either same-sign charge (SS) or opposite-sign
charge (OS) and with either same-flavour (\( ee \) or \( \mu\mu \)) or different-flavour
(\( e\mu \)) combinations.
 
The ATLAS detector is described in
\Sect{\ref{sec:detector}}. The data and simulated events used are outlined
in \Sect{\ref{sec:data-sim}}, and the event reconstruction procedure is
detailed in \Sect{\ref{sec:definitions}}. The analysis strategy is presented
in \Sect{\ref{sec:regions}}, followed by a discussion of the background estimation
in \Sect{\ref{sec:background}} and of systematic uncertainties in \Sect{\ref{sec:systematics}}.
Finally, the statistical analysis and results are presented in
\Sect{\ref{sec:results}}, followed by the conclusions in \Sect{\ref{sec:conclusion}}.
 
\FloatBarrier
 
 
\newcommand{\AtlasCoordFootnote}{
ATLAS uses a right-handed coordinate system with its origin at the nominal
interaction point (IP) in the centre of the detector and the \(z\)-axis
along the beam pipe. The \(x\)-axis points from the IP to the centre of the
LHC ring, and the \(y\)-axis points upwards. Polar coordinates
\((r,\phi)\) are used in the transverse plane, \phi being the azimuthal
angle around the \(z\)-axis. The pseudorapidity is defined in terms of the
polar angle \( \theta \) as \( \eta = -\ln \tan(\theta/2) \).
Angular distance is measured in units of
\( \Delta R \equiv \sqrt{{(\Delta\eta)}^{2} + {(\Delta\phi)}^{2}} \).}
 
\section{ATLAS detector}
\label{sec:detector}
 
The ATLAS detector~\cite{PERF-2007-01} at the LHC is a multipurpose particle
detector with a forward--backward symmetric cylindrical geometry\footnote{\AtlasCoordFootnote} and a nearly  
\( 4\pi \) coverage in solid angle around the collision point. It consists of
an inner tracking detector surrounded by a thin superconducting solenoid,
electromagnetic and hadronic calorimeters, and a muon spectrometer incorporating
three large superconducting toroidal magnets.

The inner detector is immersed in a \SI{2}{\tesla} axial magnetic
field and provides charged-particle tracking in the
range \( \abseta < 2.5 \).
The high-granularity silicon pixel detector covers the vertex region and typically
provides four measurements per track, the first hit being normally in the insertable
B-layer, which was installed before Run~2~\cite{ATLAS-TDR-2010-19,PIX-2018-001}.
It is followed by the silicon microstrip tracker which usually provides four measurements per track.
These silicon detectors are complemented by the transition radiation tracker (TRT),
which enables radially extended track reconstruction up to \(\abseta = 2.0\).
The TRT also provides electron-identification information based on the fraction
of hits (typically 30 in total) above a higher energy-deposit threshold corresponding
to that of transition radiation.
 
The calorimeter system covers the pseudorapidity range \( \abseta < 4.9 \).
Within the region \( \abseta < 3.2 \), electromagnetic calorimetry is provided
by barrel and endcap high-granularity lead/liquid-argon (LAr) sampling
calorimeters, with an additional thin LAr presampler covering \( \abseta < 1.8 \)
to correct for energy loss in material upstream of the calorimeters.
Hadronic calorimetry is provided by the steel/scintillator-tile calorimeter,
segmented into three barrel structures within \( \abseta < 1.7 \), and two
copper/LAr hadronic endcap calorimeters. The solid angle coverage is completed
with forward copper/LAr and tungsten/LAr calorimeter modules optimised
for electromagnetic and hadronic measurements, respectively.
 
The muon spectrometer comprises separate trigger and high-precision
tracking chambers which measure the deflection of muons in a magnetic field
generated by the superconducting air-core toroids. The field integral of the toroids
ranges between \num{2.0} and \SI{6.0}{\tesla\metre} across most of the muon
spectrometer.
A set of precision chambers covers the region \( \abseta < 2.7 \) with three
layers of monitored drift tubes, complemented by cathode-strip chambers in the
forward region, where the background is highest. The muon trigger system covers
the range \( \abseta < 2.4 \) with resistive-plate chambers in the barrel
and thin-gap chambers in the endcap regions.
 
A two-level trigger system is used to select interesting events~\cite{TRIG-2016-01}.
The first-level trigger is implemented in hardware and uses a subset of detector
information to reduce the event rate to at most \SI{100}{\kHz}.
This is followed by a software-based trigger which further reduces the event rate
to approximately \SI{1}{\kHz}.
 
\section{Data and simulated events}
\label{sec:data-sim}
 
The data used in this analysis correspond to proton--proton collisions 
with proton bunches colliding every \SI{25}{\ns}.
Data quality criteria are applied to ensure that considered events were recorded with
stable beam conditions and with all relevant sub-detector systems operational.
After data quality requirements, the samples used for this analysis correspond
to \SI{3.2}{\ifb}, \SI{33.0}{\ifb}, \SI{44.3}{\ifb}, and \SI{58.5}{\ifb}
of integrated luminosity recorded in 2015, 2016, 2017, and 2018, respectively,
for a total of \SI{139}{\ifb}. In the four years of data taking,
the average number of interactions per bunch crossing is 13, 25, 38, and 36,
respectively. The uncertainty in the combined 2015--2018 integrated luminosity is
\SI{1.7}{\%}~\cite{ATLAS-CONF-2019-021}, obtained using the LUCID-2
detector~\cite{LUCID2} for the primary luminosity measurements.
 
Samples of signal and background processes were simulated using several
Monte Carlo (MC) generators. The signal considered in the simplified
\typeIIIseesaw model is implemented in the \MGMCatNLO~\cite{Alwall:2014hca}
generator at leading order (\LO) using \textsc{FeynRules}~\cite{Alloul:2013bka}.
For the simulated signal production, \MGMCatNLO was interfaced to
\PYTHIAV{8.212}~\cite{Sjostrand:2014zea} for parton showering.
The \aonefour set of tuned parameters~\cite{ATL-PHYS-PUB-2014-021} was used for the parton shower.
The \nnpdflo~\cite{Ball:2014uwa} parton distribution function (PDF) set was used
in the matrix element calculation, and the \nnpdftwo~\cite{Ball:2012cx} was used
in the parton shower.
The signal cross-section and its uncertainty are calculated separately at next-to-leading-order (NLO) plus
next-to-leading-logarithmic (NLL) accuracy, assuming that the heavy leptons,
\Lpm and \Nz, are \( \text{SU}(2) \) triplet fermions~\cite{Fuks:2012qx,Fuks:2013vua}.
Simulated SM background samples include top quark pair (\ttbar) and
diboson (\( VV,~V = \Zboson, \Wboson \)) production processes, which are the dominant ones, as well as
\DY (\( \qqbar \rightarrow \Zboson/\gam^{*} \rightarrow \ellell \, (\ell=e,\mu,\tau) \)),
triboson (\( VVV \)), single top quark, and rare top quark (\( \ttbar + V \))
production processes. The non-dominant processes are grouped into the `other'
background category throughout this paper.
 
The generators used in the MC sample production and
the cross-section calculations used for MC sample normalisation are provided in
\Tab{\ref{tab:MC}}. In order to minimise the theoretical uncertainties, the normalisation
of the  MC samples modelling dominant backgrounds is considered as a free parameter in the final likelihood fit,
as described in \Sect{\ref{sec:results}}.
 
\begin{table}[tbp]
\caption{The event generator, parton shower generator, cross-section normalisation,
and PDF set used for the matrix element (ME) calculation are shown for each simulated
signal and background event sample.
The generator cross-section is used where not specifically stated otherwise.}
\label{tab:MC}
\begin{center}
{
\small
\begin{tabular}{l l c c c c}
\toprule
Physics process & Event generator & ME PDF set & Cross-section & Parton shower \\
&                 &            & normalisation &     generator\\
\midrule
 
\multirowcell{2}[0pt][l]{\Nz/\Lpm} & \multirowcell{2}[0pt][l]{\MGMCatNLO}  & \multirowcell{2}{\nnpdflo}  & \multirowcell{2}{NLO} & \multirowcell{2}{\PYTHIAV{8.212} \\ \& \EVTGEN~1.6.0} \\
& & & & \\
 
\midrule
 
Top physics & & & & \\
\ttbar & \POWHEGBOXV{v2}  & \nnpdfnlo &
NNLO+NNLL  & \multirowcell{3}{\PYTHIAV{8.230} \\ \& \EVTGEN~1.6.0} \\
 
Single \( t \) & \POWHEGBOXV{v2} & \nnpdfnlo &
NLO & \\
 
\( \ttbar + V \) & \POWHEGBOXV{v2} & \nnpdfnlo &
NLO &  \\
 
\midrule
 
Multiboson & & & \\
\( \Zboson\Zboson \), \( \Wboson\Zboson \), \( \Wboson\Wboson \) & \SHERPAV{2.2.1} \& 2.2.2  & \nnpdfnnlo &
NLO & \SHERPA  \\
 
\( VVV \) & \SHERPAV{2.2.2} & \nnpdfnnlo &
NLO & \SHERPA \\
 
\midrule
 
\DY & & & & & \\
\( \Zboson/\gam^{*}\rightarrow \ell^+ \ell^- \) &
\SHERPAV{2.2.1} & \nnpdfnnlo &
NNLO  & \SHERPA \\
 
\bottomrule
\end{tabular}
}
\end{center}
\end{table}

The production of \ttbar\ events was modelled using the
\POWHEGBOXV{v2} generator~\cite{Frixione:2007nw,Nason:2004rx,Frixione:2007vw,Alioli:2010xd},
which provides matrix elements at next-to-leading order~(NLO) in the
strong coupling constant \alphas with the \nnpdfnlo
PDF.\@ The \hdamp parameter in \POWHEG controls the
matching between the matrix element and the parton shower, and effectively
regulates the high-\pt\ radiation against which the \ttbar\ system recoils.
It was set to \SI{1.5}{\mtop}~\cite{ATL-PHYS-PUB-2016-020},
using a top quark mass of \( \mtop = \SI{172.5}{\GeV} \).
The renormalisation and factorisation scale was
set to {\small \( \sqrt{m_\textrm{top}^2 + p_\textrm{T, top}^2} \)}, where \(p_\textrm{T, top} \) is the transverse momentum of the simulated top quark in the event.
The events were interfaced with \PYTHIAV{8.230}~\cite{Sjostrand:2014zea}
for the parton shower and hadronisation, using the \aonefour set of tuned
parameters and the \nnpdftwo set of PDFs.
The decays of bottom and charm hadrons were simulated using
\EVTGEN v1.6.0~\cite{Lange:2001uf}.
The \ttbar sample is initially normalised to the cross-section prediction at
next-to-next-to-leading order (NNLO) in QCD including the re-summation of
next-to-next-to-leading logarithmic (NNLL) soft-gluon terms calculated using
\textsc{Top++2.0}~\cite{Beneke:2011mq,Cacciari:2011hy,Baernreuther:2012ws,Czakon:2012zr,Czakon:2012pz,Czakon:2013goa,Czakon:2011xx},
while the final yield is extracted from the data, as is described in \Sect{\ref{sec:regions}}.
For \pp collisions at a centre-of-mass energy of \( \rts = \SI{13}{\TeV} \),  
this cross-section corresponds to \( \sigma{(\ttbar)}_{\textrm{NNLO+NNLL}} = \mathrm{832\pm51~pb} \).
 
Samples of diboson final states (\(VV\)) were simulated with the
\SHERPAV{v2.2.1} or v2.2.2 generator~\cite{Bothmann:2019yzt}, as detailed in \Tab{\ref{tab:MC}},
including off-shell effects and Higgs boson contributions,
where appropriate. Fully leptonic final states and semileptonic final states,
where one boson decays leptonically and the other hadronically, were generated
using matrix elements at NLO accuracy in QCD for up to one additional parton
and at LO accuracy for up to three additional parton emissions.
Samples for the loop-induced processes \( gg \to VV \) were
generated using LO-accurate matrix elements for emission of up to one additional parton
for both the fully leptonic and semileptonic final states.
Electroweak production of a diboson pair in association with two jets (\(VVjj\)) is
also accurate up to leading order.
The matrix element calculations were matched
and merged with the \SHERPA parton shower based on Catani--Seymour  
dipole factorisation~\cite{Gleisberg:2008fv,Schumann:2007mg} using the MEPS@NLO prescription~\cite{Hoeche:2011fd,Hoeche:2012yf,Catani:2001cc,Hoeche:2009rj}.
QCD corrections were provided by the \OPENLOOPSV{library}~\cite{Cascioli:2011va,Denner:2016kdg-fixed}.
The \nnpdfnnlo set of PDFs was used~\cite{Ball:2014uwa}, along with the
dedicated set of tuned parton-shower parameters developed by the
\SHERPA authors.
 
Other background processes, such as single-top production, \( \ttbar + V \) processes and the
\DY  \( \Zboson/\gam^{*}\rightarrow \ee/\mumu/\tautau \) process,
do not contribute significantly to the regions defined in this analysis and
are outlined only in \Tab{\ref{tab:MC}}.
 
For all samples, a full simulation of the ATLAS detector response~\cite{SOFT-2010-01}
was performed using the \GEANT4 toolkit~\cite{Agostinelli:2002hh}.
The effect of multiple interactions in the same and neighbouring bunch
crossings (pile-up) was modelled by overlaying the original hard-scattering
event with simulated inelastic \(pp\) events generated with \PYTHIAV{8.186}
using the \nnpdftwo set of PDFs and the \athree set of tuned
parameters~\cite{ATL-PHYS-PUB-2016-017}.
The MC events are weighted to reproduce the
distribution of the average number of interactions per bunch crossing
(\( \left<\mu \right> \)) observed in the data. The \( \left<\mu \right> \)
value in MC is rescaled by a factor of \( 1.03 \pm 0.07 \) to improve
the level of agreement between data and simulation in the visible inelastic
\( pp \) cross-section~\cite{STDM-2015-05}.
 
 
\section{Event reconstruction}
\label{sec:definitions}
 
After the application of beam, detector and data-quality requirements, events
containing jets failing to satisfy the quality criteria described in \Refn{\cite{ATLAS-CONF-2015-029}}
are rejected to suppress events with large calorimeter noise or non-collision backgrounds.
To further reject non-collision backgrounds originating from cosmic rays and beam-halo events,
at least one reconstructed collision vertex is required with at least two associated
tracks with transverse momentum (\pt) greater than \SI{500}{\MeV}.
The \pp vertex candidate with the highest sum of the squared transverse momenta
of all associated tracks is identified as the primary vertex.

Electron candidates are reconstructed from energy deposits in the electromagnetic
calorimeter associated with a charged-particle track measured in the inner
detector. The electron candidates are required to pass the \textit{Tight}
likelihood-based identification selection~\cite{PERF-2013-05,PERF-2017-01,ATLAS-CONF-2016-024},
to have transverse momentum \( \pt > \SI{40}{\GeV} \) and to be in the fiducial
volume of the inner detector, \( \abseta < 2.47 \). The transition region
between the barrel and endcap calorimeters (\( 1.37 < \abseta < 1.52 \))
is excluded because it has inefficient regions due to the presence of service conduits.
The track associated with the electron candidate must have a transverse impact parameter
(\dzero) evaluated in the transverse plane at the point of closest approach of the track to the beam axis which satisfies
\( |\dzero|/\sigma(\dzero) < 5 \), where \( \sigma(\dzero) \) is the uncertainty
in \dzero. A longitudinal impact parameter value of
\( |z_{0}\sin(\theta)| < \SI{0.5}{mm} \) is also required for electrons,
where \zzero is the distance between the track and the primary vertex in the \(z\) direction and
\( \theta \) is the polar angle of the track. The combined identification
and reconstruction efficiency for \textit{Tight} electrons ranges from
\SI{58}{\%} to \SI{88}{\%} in the transverse energy (\ET) range from \SI{4.5}{\GeV} to
\SI{100}{\GeV}. In addition, the electron candidates must also satisfy the
\textit{Loose} isolation criteria, which use calorimeter-based and track-based isolation
requirements with an efficiency of about \SI{99}{\%}. Both reconstruction and isolation performance are
evaluated in \Zee decay measurements, described
in \Refn{\cite{EGAM-2018-01}}. Electron candidates are discarded if their
angular distance from a jet satisfies \( 0.2 < \DeltaR < 0.4 \).
 
Muon candidates are reconstructed by matching an inner-detector track with a track
reconstructed in the muon spectrometer. The muon candidates
are required to have \( \pt > \SI{40}{\GeV} \), \( \abseta < 2.5 \) and a transverse
impact parameter significance of \( |\dzero|/\sigma(\dzero) < 3 \).
A longitudinal impact parameter value of
\( |z_{0}\sin(\theta)| < \SI{0.5}{mm} \) is required for muons.
Muon candidates with \pt lower than \SI{300}{\GeV} are required to pass the \textit{Medium}
muon identification requirements, while for higher values of \pt a special \textit{HighPt} muon identification
is applied, which is optimised for searches in the high-\pt regime~\cite{PERF-2015-10}.
The muon candidates must also fulfil the \textit{TightTrackOnly} track-based
isolation requirements. This results in a reconstruction and identification
efficiency of over \SI{95}{\%} in the entire phase space, as measured
in \Zmm events~\cite{PERF-2015-10}. Muons within \( \Delta R = 0.4 \) of the axis of a jet
having more than three tracks and with \( \pt(\mu) / \pt(\mathrm{jet}) < 0.5 \)
are discarded. If a muon candidate overlapping with an electron leaves a sufficiently
large energy deposit in the calorimeter and shares a track reconstructed
in the inner detector with the electron, this muon candidate is also discarded.
 
Jets are reconstructed by clustering energy deposits in the calorimeter using
the \antikt algorithm~\cite{Cacciari:2008gp}  with the radius parameter equal to 0.4.
The measured jet transverse momentum is corrected for detector effects
by recalibrating to the particle energy scale before the interaction with
the detector~\cite{PERF-2016-04}. For all jets the expected average transverse energy contribution
from \pileup is corrected using a subtraction method. This is based on the level
of diffuse noise added to the event per unit area in \( \eta \)--\( \phi \) by \pileup which
is subtracted from the jet area in \( \eta \)--\( \phi \) space and a residual correction
derived from the MC simulation, as detailed in \Refns{\cite{PERF-2016-04, Cacciari:2007fd}}.
To reduce the contamination from jets from \pileup,
a `jet vertex tagger' (JVT) algorithm~\cite{ATLAS-CONF-2014-018} is used for
jets with \( \pt < \SI{60}{GeV} \) and \( \abseta < 2.4 \). It employs
a multivariate technique that relies on jet energy and tracking variables
to determine the likelihood that the jet originates from \pileup.
The \textit{Medium} JVT working point is used which has an average efficiency
of \SI{92}{\%}. Jets within \( \Delta R = 0.2 \) of an electron are discarded.
Jets within \( \Delta R = 0.2 \) of a muon and featuring fewer than three tracks
or having \( \pt(\mu) / \pt(\mathrm{jet}) > 0.5 \) are also removed.
Jets considered in this analysis are required to have \( \pt > \SI{20}{GeV} \)
and \( \abseta < 2.5 \). Jets fulfilling these kinematic criteria are
identified as containing \(b\)-hadrons and categorised as \(b\)-tagged jets
if tagged by a multivariate algorithm which uses information about the impact
parameters of inner-detector tracks matched to the jet, the presence of displaced
secondary vertices, and the reconstructed flight paths of \(b\)- and \(c\)-hadrons
inside the jet~\cite{PERF-2012-04,ATL-PHYS-PUB-2016-012,ATL-PHYS-PUB-2017-013}.
The algorithm is used at the working point providing a \(b\)-tagging efficiency
of \SI{77}{\%}, as determined in a sample of simulated \ttbar events.
This corresponds to rejection factors of approximately 134, 6 and 22 for
light-quark and gluon jets, \(c\)-jets, and hadronically-decaying
\( \tau \)-leptons, respectively.
 
Missing transverse momentum
(with magnitude \MET) is present in events with unbalanced kinematics in the
transverse plane. This originates from undetected momentum in the event,
due either to neutrinos escaping detection or to other particles falling outside
the detector acceptance, being badly reconstructed, or failing reconstruction altogether.
The \MET is calculated as the modulus of the negative vectorial sum of the \pt
of the fully calibrated and reconstructed physics objects in the event.
Jets in the forward direction are considered if they satisfy \( \pt > \SI{30}{GeV} \).
An additional `soft term' is added, accounting for all tracks in the event
that are not associated with any reconstructed object, but are associated with
the identified primary vertex. The use of this track-based soft term
is motivated by improved performance in \MET reconstruction in a high \pileup
environment~\cite{PERF-2016-07}.
 
Correction factors to account for differences in the identification and selection
efficiency, reconstructed energy and energy resolution between data and
MC simulation are applied to the selected electrons, muons and jets,
and consequently also when calculating \MET in MC simulation, as described in
\Refns{\cite{PERF-2013-05,ATLAS-CONF-2016-024,PERF-2015-10,PERF-2016-04}}.
 
\section{Analysis strategy and event selection}
\label{sec:regions}
 
All selected events, reconstructed as defined in \Sect{\ref{sec:definitions}}, are required to match the signal topology, as defined in  \Sect{\ref{sec:intro}}, whereby a separate analysis channel is used for each light-lepton flavour (\( ee \), \( e\mu \), \( \mu\mu \)) and charge combination (SS, OS), resulting in six analysis channels.
Events must pass dilepton triggers, where the \pt thresholds are optimised depending on the lepton flavour combination and the data-taking year.
Thresholds range from \SI{12}{\GeV} to \SI{22}{\GeV} for the leading lepton \pt
and from \SI{8}{\GeV} to \SI{17}{\GeV} for the \pt of the sub-leading lepton.
Compared to single-lepton triggers, the dilepton triggers have looser
identification and isolation criteria, minimising
biases on the implementation of data-driven estimation of backgrounds,
with only a marginal reduction of signal event efficiencies.

Selected events are then categorised into exclusive categories
(\textit{analysis regions}) based on different sets of requirements for reconstructed
objects. These regions are grouped according to their purpose into signal regions, control regions, and validation regions.
Signal regions (SRs) are defined to achieve the best sensitivity to the targeted models with an enhanced signal-to-background ratio and are used to compare data with each signal hypothesis plus the expected background
using the statistical methodology detailed
in \Sect{\ref{sec:results}}. The purpose of control regions (CRs) and  validation regions (VRs) is to evaluate and validate the background contamination in the SRs. The CRs and VRs are thus constructed with the purpose of minimising the signal presence while maximising the contributions of distinct types of background. Consequently, the CRs and VRs are constructed to be kinematically close to the SRs, which is done by modifying one or more kinematic requirements while also ensuring no overlap with SRs and among themselves. In CRs the background prediction is fit to data by assigning a floating parameter to the background normalisation of the main process. In VRs the background estimation methods are validated by comparing the background model with data.
 
While optimising the SR definition to maximise the signal significance, the data are not looked at (blinded), until the background modelling has been validated. Likewise, as a part of the same strategy,
while validating the background predictions, the data are compared with the background estimates
only in the CR and VR, which are designed to have negligible signal contamination.
 
All regions used in this analysis, with the corresponding selection criteria,
are summarised in \Tab{\ref{tab:regions}} and described below.
 
\begin{table}[bp]
\caption{A summary of all analysis regions defined in the text.
The region definitions are the same for all analysis channel flavour
combinations.}
\label{tab:regions}
\begin{center}
\setlength\tabcolsep{2.75pt}
{
\small
\begin{tabular}{ l | c c c | c c c } 
\toprule
& \multicolumn{3}{c |}{\textbf{OS}
(\( \ell^+\ell^-  = e^+e^-, e^\pm\mu^\mp, \mu^+\mu^- \))}
& \multicolumn{3}{c}{\textbf{SS}
(\( \ell^\pm\ell^\pm = e^\pm e^\pm, e^\pm \mu^\pm, \mu^\pm\mu^\pm \))}
\\
& Top CR                           & \mjj VR                        & SR               & Diboson CR                    & \mjj VR                       & SR               \\
\midrule
\( N(\text{jet}) \)                         & \( \geq 2 \)                     & \( \geq 2 \)                   & \( \geq 2 \)     & \( \geq 2 \)                  & \( \geq 2 \)                  & \( \geq 2 \)     \\
\( N(\bjet) \)                              & \(   {\geq 2} \)                 & 0                              & 0                & 0                             & 0                             & 0                \\
\( \mjj \) \, [\GeV]                        & \( (60, 100) \)                  & \( (35, 60) \cup (100, 125) \) & \( (60, 100) \)  & \( (0, 60) \cup (100, 300) \) & \( (0, 60) \cup (100, 300) \) & \( (60, 100) \)  \\
\( \mll \) \, [\GeV]                        & \( \geq 110 \)                   & \( \geq 110 \)                 & \( \geq 110 \)   & \( \geq 100 \)                & \( \geq 100 \)                & \( \geq 100 \)   \\
\metsig                                     & \( \geq 5 \)                     & \( \geq 10 \)                  & \( \geq 10 \)    & \( \geq 5 \)                  & \( \geq 5 \)                  & \(  \geq 7.5 \)  \\
\( \Delta\phi{(\met, \ell)}_\mathrm{min} \) & ---                              & ---                            & \( \geq 1 \)     & ---                           & ---                           & ---              \\ 
\( \pt(jj) \) \, [\GeV]                     & ---                              & ---                            & \(   \geq 100 \) & ---                           & ---                           & \( \geq 60 \)    \\
\( \pt(\ell\ell) \) \, [\GeV]               & ---                              & ---                            & \(   \geq 100 \) & ---                           & ---                           & \(   \geq 100 \) \\
\( \HT+\met \) \, [\GeV]                    & \( \geq 300 \)                   & \( \geq 300 \)                 & \( \geq300 \)    & \( (300, 500) \)              & \(   { \geq 500 } \)          & \( \geq 300 \)   \\
\bottomrule
\end{tabular}
}
\end{center}
\end{table}
 
As the signal process contains neutrinos in the final state, one of the most
important selection criteria is based on the \MET significance
\metsig~\cite{PERF-2016-07}. The value of \metsig is calculated using
a maximum-likelihood ratio method which considers the direction of the \MET and the calibrated
objects as well as their respective resolutions. The SR selection criteria are
optimised to maximise the analysis sensitivity. This results in different
selection values for the OS and SS channels due to different background compositions and associated event topologies. Consequently, the \MET significance
selection value is \(\metsig > 10\) for the OS channels and \(\metsig > 7.5\) for
the SS channels.
 
At least two jets with \( \pt > \SI{20}{\GeV} \) and \( \abseta < 2.5 \) are
required for each region. A \bjet veto is applied to all the jets in SR events
to suppress background from SM processes involving top quarks.
For each signal event, the dijet invariant mass (\mjj) of the two
highest-\pt jets is expected to be close to the \Wboson mass. The dijet
invariant mass is thus required to be in the window
\( \SI{60}{\GeV} < \mjj < \SI{100}{\GeV} \).
 
In the SR definition, a lower bound on the invariant mass of the lepton pair (\mll)
is introduced and is found to give optimal results at \SI{110}{\GeV} (\SI{100}{\GeV}) in the OS (SS) regions.
This choice aims to remove the \DY events around the \Zee peak, where the electron charge might also have been misidentified.
 
Furthermore, in the SR definition for the OS analysis channels, the azimuthal angle \( \Delta\phi{(\met, \ell)}_\mathrm{min} \) 
between the directions of \met and the closest lepton has been shown to have
good separation power between signal and background events, especially reducing the SR background contamination from \ttbar and diboson events.
This is explained by difference between the sources of the measured \met for signal
and background, because the latter tends to have a spurious component due to
misreconstructed jets. After optimisation,  a requirement of
\( \Delta\phi{(\met, \ell)}_\mathrm{min} > 1 \) is used. In the SR definition for the SS analysis channels, this selection criterion is not applied, because other criteria already reduce the \ttbar and diboson background contamination, and this criterion is not very effective in reducing the fake-background contribution, whose contribution is significantly larger for the SS analysis channels.  
 
To further increase the expected signal significance, additional selection
criteria are introduced: the dijet transverse momentum must fulfil the condition \( \pt(jj) > \)~\SI{100}{\GeV} (\SI{60}{\GeV}) for the OS (SS) regions,
while the dilepton transverse momentum must satisfy \( \pt(\ell\ell) > \)~\SI{100}{\GeV}
for both OS and SS events. These selection requirements exploit
the boosted decay topology of object pairs, which would be expected in the presence
of heavy leptons.

Due to the high masses of the heavy leptons in the decay chain, the signal
process is expected to contain high-\pt leptons, jets and
neutrinos. Since the optimal measure for the high-\pt activity of the neutrinos escaping the detector
is the \met,  and an estimator of the event activity for the measured objects is the scalar sum of the transverse momenta of selected leptons and jets \HT, a combined \( \HT+\met > \SI{300}{\GeV} \) selection criterion is applied in all analysis
regions. The \HT+\met is also the observable used in the final likelihood fit,
described in \Sect{\ref{sec:results}}.
The final signal selection efficiency relative to the events with at least two leptons
is about \SI{1.2}{\%} for masses between \SI{500}{\GeV} and \SI{900}{\GeV} and drops below \SI{1}{\%} for higher mass points.
 
The dominant background contribution (almost \SI{50}{\%} of the total) in the OS channels is \ttbar
production in which the two \Wboson bosons in the final state decay leptonically.
In the SS channels the \ttbar events originate from charge misidentification
and are at the level of \SI{25}{\%}.
To estimate the contribution from the \ttbar decays, an OS control region enriched
in top-quark events is defined (Top CR); no dedicated SS CR is used.
In this region, the SR selection is modified
by requiring the number of \(b\)-tagged jets to be \( N(\bjet) \geq~2\).
To increase the statistical significance, all requirements on the transverse momenta,
\( \pt(jj) \) and \( \pt(\ell\ell) \), as well as the azimuthal angle, \( \Delta\phi{(\met, \ell)}_\mathrm{min} \), are omitted, and the  \metsig selection is relaxed to \( \metsig > 5 \).  
The normalisation of the \ttbar MC sample is a free parameter
in the simultaneous likelihood fit across all CRs, as described in
\Sect{\ref{sec:results}}.
 
\begin{figure}[tpb]
\centering
\subfloat[]{
\includegraphics[width=0.48\textwidth]{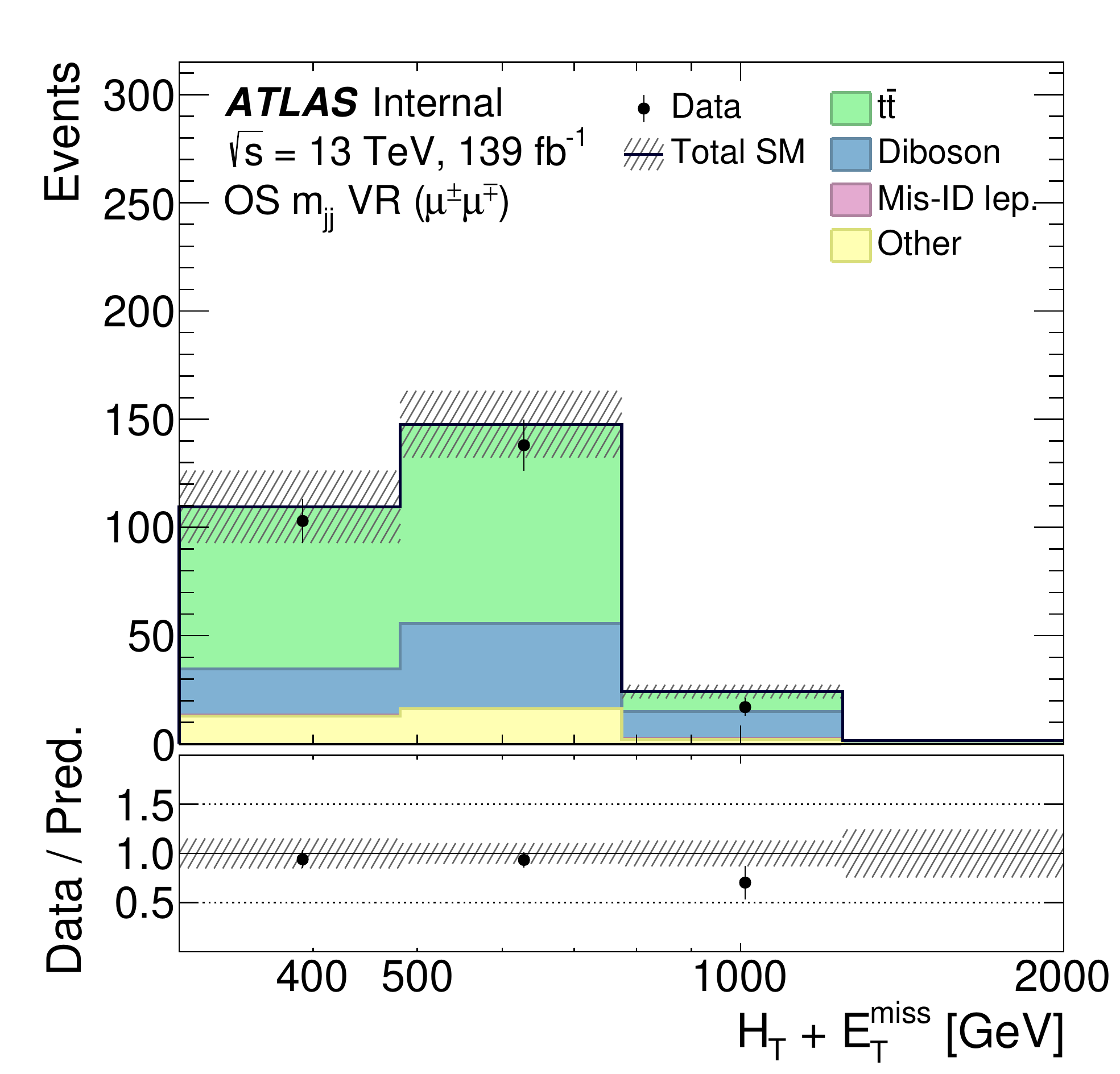}
\label{fig:Repr_VR-1}
}
\subfloat[]{
\includegraphics[width=0.48\textwidth]{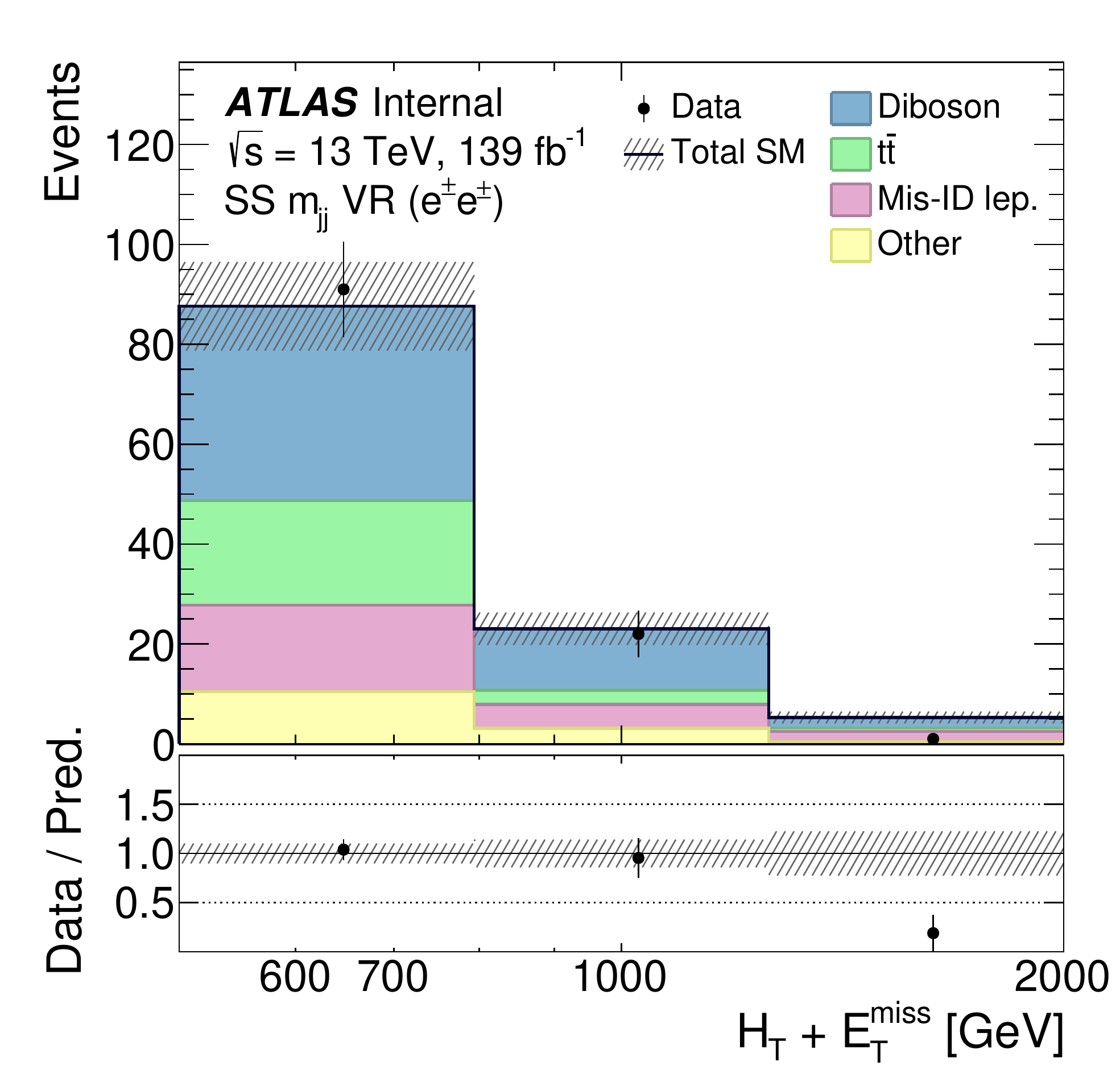}
\label{fig:Repr_VR-2}
}
\caption{Post-fit distributions of the \( \HT + \met \) variable for data and SM
background predictions in two validation regions, namely
\protect\subref{fig:Repr_VR-1}
in the OS \( \mu\mu \)~\( \mjj \) VR and
\protect\subref{fig:Repr_VR-2}
in the SS \( ee \)~\( \mjj \) VR.\@
The hatched bands include all systematic uncertainties post-fit
with the correlations between various sources taken into account.
Errors on data are statistical only.
The lower panel shows the ratio of the observed data to the estimated
SM background.
}
\label{fig:Repr_VR}
\end{figure}

To obtain validation and control regions with a large fraction of diboson events, the dijet mass \mjj selection is inverted relative to the SR definition. Since the \mjj selection in the SR requires a pair of jets to match the \Wboson mass window, the inverted selection defines a kinematic sideband, with
negligible signal contamination but still selecting events kinematically similar to those in the SR.\@ In addition, all requirements on the transverse momenta,
\( \pt(jj) \) and \( \pt(\ell\ell) \), as well as the azimuthal angle, \( \Delta\phi{(\met, \ell)}_\mathrm{min} \), are omitted, in order to ensure that there are enough events in the CRs. 
This selection defines the \mjj VR for the OS analysis channels, while for the SS analysis channels the
\metsig selection is additionally relaxed to \metsig \( > 5 \).  The SS Diboson CR is then defined by introducing an additional requirement \( \SI{300}{\GeV} < \HT+\met < \SI{500}{\GeV} \), while
the remaining kinematic region with \(\HT+\met > \SI{500}{\GeV} \) is used as the SS \mjj VR.\@
The diboson background contributes only slightly less (\SI{45}{\%}) than the \ttbar background to the contamination in the OS SR and more than half (\SI{55}{\%}) of the total background in the SS SR.\@
The normalisation of the diboson MC sample used in the analysis is estimated
in the SS Diboson CR for both OS and SS events, as the OS and SS diboson
cross-sections are found to be the same within the systematics uncertainties.
This is achieved by using the normalisation value of the diboson MC sample as a free parameter in the simultaneous likelihood fit across all CRs, as described in \Sect{\ref{sec:results}}.
Two representative post-fit distributions of the \( \HT + \met \) variable in the VRs are shown in \Fig{\ref{fig:Repr_VR}}.
Good agreement between measured data and predictions is observed within uncertainties.
 
\section{Background composition and estimation}
\label{sec:background}
 
The final-state topologies of the six analysis channels have different background
compositions. Background estimation techniques, implemented consistently among
the channels, are discussed in this section. In general, the expected number of
background events and the associated kinematic distributions are derived with
a mixture of data-driven methods and simulation.
 
In the analysis, one considers two conceptually different background categories according to their source, namely irreducible and reducible backgrounds.
Irreducible background sources are SM processes producing the same prompt final-state lepton pairs
as does the signal, with the dominant contributions in this analysis coming from \ttbar and diboson production. Irreducible-background predictions are obtained from simulated event samples, described
in \Sect{\ref{sec:data-sim}}.
To avoid double-counting between the irreducible background predictions, which are derived from MC, and the data-driven reducible background predictions, a dedicated algorithm is introduced. Irreducible MC events are only used if simulated leptons originating in the hard process (prompt leptons) can be associated to their reconstructed lepton counterpart.
 
The reducible background sources are the processes where the leptons reconstructed in the analysis originate from either non-prompt leptons or other particles and thus the selected events contain at least one fake/non-prompt electron or muon. In addition, events where a lepton charge is misidentified (allowing the \ttbar process to contribute to the background in the SS analysis channels)
are also considered as a reducible background category.
 
The reducible background due to charge
misidentification is estimated in SS analysis channels that contain electrons.
The effect of muon charge misidentification is found to be negligible.
The modelling of charge
misidentification in simulation deviates from that in data due to the complexity of the processes
involved, and it relies strongly on a precise description of the detector
material.
Consequently, charge reconstruction correction factors (scale factors),
applied to the simulated background events to compensate for the differences relative to data,
are derived by comparing the charge misidentification probability measured in data with
the one in simulation. The charge misidentification probability is extracted
by performing a likelihood fit on a dedicated \Zee data sample, as described in \Refn{\cite{EGAM-2018-01}}.
These scale factors are then applied to the simulated background events to compensate for the differences.

The `fake-background' sources, non-isolated, non-prompt electrons and muons, are produced by secondary decays
of light- or heavy-flavour mesons into light leptons embedded within jets. Although the \(b\)-jet veto significantly reduces the number of fake leptons from heavy-flavour decays, a fraction of these events still passes the analysis selection.
For electrons, another significant component of fakes arises from photon conversions, as well as
from jets that are reconstructed as electrons.
MC samples are not used to estimate these background sources because the simulation
of jet production and hadronisation has large intrinsic uncertainties.
Instead, the fake-lepton background (namely, background from  misidentified or non-prompt leptons) is estimated with a data-driven approach, known as the `fake factor' method, as described in \Refn{\cite{EXOT-2012-25}}. This method relies on determining the probability, or fake factor (\(F\)), for a fake lepton to be identified as a tight (\(T\)) lepton, corresponding to the default lepton selection. The fake factor is measured in dedicated fake-enriched regions, where events must pass single-lepton triggers, contain no \(b\)-jets and have only one reconstructed lepton that satisfies either the tight selection criteria or a relaxed (loose, \(L\)) selection. Loose electrons will pass the \textit{Loose} identification requirements but fail either the \textit{Tight} identification or the isolation requirements. Loose muons, on the other hand, will pass the \textit{Medium} identification requirements but fail isolation. Electron and muon \textit{fake factors} \(F\) are then defined as the ratio of the number of tight leptons to the number of loose leptons and are parameterised as functions of \pt and \eta. The fake-lepton background (containing at least one fake lepton) is then estimated in the SR by applying \(F\) as a normalisation correction to relevant distributions in a template region which has the same selection criteria as the SR except that at least one of the two leptons must pass the loose selection but fail the tight one.
 
The number of events in the SR that contain at least one fake lepton,
\( N^{\mathrm{fake}} \), is estimated from the adjacent regions, which can be labelled by the
lepton pair passing/failing the nominal requirements as  (\(TL\), \(LT\), \(LL\)), where
the first lepton is leading and the second is sub-leading in \pt.
Data events are weighted with fake factors according to the loose-lepton multiplicity of the region:
\begin{equation*}
N^{\mathrm{fake}} =
{\left[ F(N_{TL} + N_{LT}) - F^2N_{LL} \right]}_{\mathrm{data}}
- {\left[ F(N_{TL} + N_{LT}) - F^2N_{LL} \right]}^{\mathrm{prompt~\ell~only}}_{\mathrm{MC}},
\end{equation*}
with \(N_{TL}\), \(N_{LT}\), \(N_{LL}\) denoting the number of events in the
corresponding adjacent region. The prompt-lepton contribution is subtracted
using the irreducible-background MC samples to account for the prompt-lepton
contamination in the adjacent regions.
The \mjj VR, as defined in \Tab{\ref{tab:regions}}, is used to verify
the data-driven fake-lepton estimate.
 
The experimental systematic uncertainty of the data-driven estimate of the
fake-lepton background is evaluated by rederiving the fake factor while varying
the relevant systematic uncertainty sources by their uncertainties.
These account for variation in the underlying
composition (i.e.\@ relative contributions of different background processes)
of fake backgrounds across the regions analysed. Furthermore, the uncertainty
in the normalisation of the simulated irreducible events is accounted for
in the background estimation procedure when the prompt-lepton contribution
gets subtracted.
The total uncertainty in the fake-lepton background varies between \SI{10}{\%}
and \SI{20}{\%} across the analysed kinematic range.
 
\section{Systematic uncertainties}
\label{sec:systematics}
 
Experimental and theoretical systematic uncertainty sources affecting both
the background and signal predictions are accounted for in the analysis.
The impact of systematic uncertainties on the total event yields as well
as the changes in the shapes of kinematic distributions are taken into account
when performing the statistical analysis in both the SRs and CRs,
as described in \Sect{\ref{sec:results}}.
 
Uncertainties in the predicted background yields and kinematic
distributions of the MC samples (\ttbar and diboson production) arise from
variations in parton-shower parameters, uncertainties in the QCD
renormalisation/factorisation scales, \( \alphas(\mZ) \) uncertainty, and
uncertainties due to PDFs used.
All uncertainties listed are calculated using the PDF4LHC prescription~\cite{Butterworth:2015oua}.
Since the yield  of the \ttbar and diboson backgrounds is derived from
the likelihood fit to the data in the CRs, these systematic variations
contribute by changing the shapes of the background predictions used
in the likelihood fit in the CRs and SRs.
 
The effect of the uncertainty in the strong coupling constant \( \alphas(\mZ) = 0.118 \) is assessed
by variations of \( \pm 0.001 \).
Uncertainties from missing higher orders are evaluated by using seven variations
of the QCD factorisation and renormalisation scales in the matrix elements
up to a factor of two~\cite{Bothmann:2016nao}.
Uncertainties in the nominal PDF set, used
in the MC generation, are evaluated from 100
variations using the \texttt{LHAPDF}6 toolkit~\cite{Buckley:2014ana}.
Additional uncertainties in the \ttbar acceptance due to the selection of the PDF set are calculated
using the central values of the \mstw \SI{68}{\%} CL NNLO~\cite{Martin:2009iq,Martin:2009bu},
\ctten NNLO~\cite{Lai:2010vv,Gao:2013xoa} and \nnpdftwo 5f FFN~\cite{Ball:2012cx} PDF sets.
In \SHERPA diboson samples central values of the \ctfourteen~\cite{Dulat:2015mca}
and \mmht~\cite{Harland-Lang:2014zoa} PDF sets are used.
 
The uncertainty due to initial-state-radiation (ISR) modelling is estimated by
comparing the nominal \ttbar\ sample with two additional
samples, produced by simultaneously varying the factorisation and renormalisation scales by a factor
of two~\cite{ATL-PHYS-PUB-2017-007-fixed}. Similarly, the impact of
final-state-radiation (FSR) modelling is evaluated by a reweighting procedure, using a variation of the FSR parton shower, in which  the renormalisation scale for QCD emission is
varied  by factors of 0.5 and 2.0.
The impact of the parton shower and hadronisation model is evaluated
by comparing the nominal generator set-up with a sample showered with
\HerwigV{7.13}~\cite{Bahr:2008pv,Bellm:2015jjp}, using the \HerwigV{7.1}~default set
of tuned parameters~\cite{Bellm:2015jjp} and the \mmht~PDF set.
To assess the uncertainty in the matching of NLO matrix elements and
the parton shower, the \POWHEG sample is compared with a sample of events
generated with \MGMCatNLOV{v2.6.0} interfaced with \PYTHIAV{8.230}.
 
For signal samples, only uncertainties in the QCD scales and PDFs used are considered.
As signal normalisation is the unknown parameter being measured,
only the acceptance effect on the limits is taken into account.
 
A significant contribution to the total background uncertainty arises from
the statistical uncertainty in the data and MC samples.
The corresponding data statistical uncertainty in the signal regions varies
from \SI{29}{\%} to \SI{47}{\%}, and the MC statistical uncertainty
from \SI{7}{\%} to \SI{10}{\%}, depending on the region considered.
Additionally the \SI{1.7}{\%} uncertainty in the combined 2015--2018
integrated luminosity, as defined in \Sect{\ref{sec:data-sim}}, is taken into account.
 
Experimental systematic uncertainties due to different lepton reconstruction,
identification, charge identification, isolation, and trigger efficiencies
in data compared to those in simulation are also taken into
account~\cite{ATLAS-CONF-2016-024,PERF-2015-10}, as are uncertainties
in absolute lepton energy calibration~\cite{PERF-2013-05}.
Likewise, experimental systematic uncertainties due to different reconstruction
and \btag efficiencies for reconstructed jets in data and simulation are
also taken into account~\cite{PERF-2012-04,ATL-PHYS-PUB-2016-012,ATL-PHYS-PUB-2017-013},
as well as absolute jet and \met energy scale and resolution uncertainties~\cite{PERF-2016-04,PERF-2016-07}.
The uncertainty in the \pileup simulation, derived from comparison of data with simulation, is also taken
into account~\cite{ATLAS-CONF-2014-018}. The uncertainty in the data-driven
estimate of the fake-lepton background is evaluated by varying the fake factor
as described in \Sect{\ref{sec:background}}.
All experimental systematic uncertainties discussed here affect the signal
samples as well as the background.
 
\section{Statistical analysis and results}
\label{sec:results}
 
The statistical analysis package \textsc{HistFitter}~\cite{Baak:2014wma}
is used to implement a binned maximum-likelihood fit in all control and signal regions,
described in  \Sect{\ref{sec:regions}}, to obtain the numbers
of signal and background events. Validation regions are used to cross-check the fit
results but are not included in the fit itself.
For the likelihood fit, the CRs and SRs
are binned in the \( \HT+\met \) variable in bins uniformly defined in \( \log(\HT+\met) \)
in the range \( \SI{300}{\GeV} < (\HT+\met) < \SI{2}{\TeV} \), where the last bin
also includes any overflow values. The SRs have six bins, and the Top CRs and
Diboson CRs each have two bins. The VRs have four and three bins for OS and SS events, respectively.
 
The binned likelihood function used in the final maximum-likelihood fit  is a product of  Poisson probability density functions that compares the observed number of events with each signal hypothesis plus the expected background, and assumes Gaussian distributions to constrain the nuisance parameters associated with the systematic uncertainties for each bin in the fitted distributions. The systematic uncertainties of the sources listed in \Sect{\ref{sec:systematics}} are used to compute the uncertainties of all the bins in the fit and are then mapped to the nuisance parameters. The widths of the Gaussian distributions correspond to the magnitudes of these uncertainties. Poisson distributions are used for modelling MC simulation statistical uncertainties.
 
The final normalisations of the \ttbar and
diboson MC samples are not taken from MC calculations but are derived in the
simultaneous likelihood fit to the data in the dedicated Top and Diboson CRs,
as introduced in \Sect{\ref{sec:regions}}. Consequently, additional free parameters
in the likelihood are introduced for the \ttbar and diboson
background contributions, to fit their yields in the Top and Diboson
control regions, respectively.  Fitting the yields of the largest backgrounds
reduces the systematic uncertainty in the predicted yield from SM sources.
The fitted normalisations are compatible with their SM predictions within
the uncertainties.

\begin{figure}[htpb]
\centering
\subfloat[]{
\includegraphics[width=0.48\textwidth]{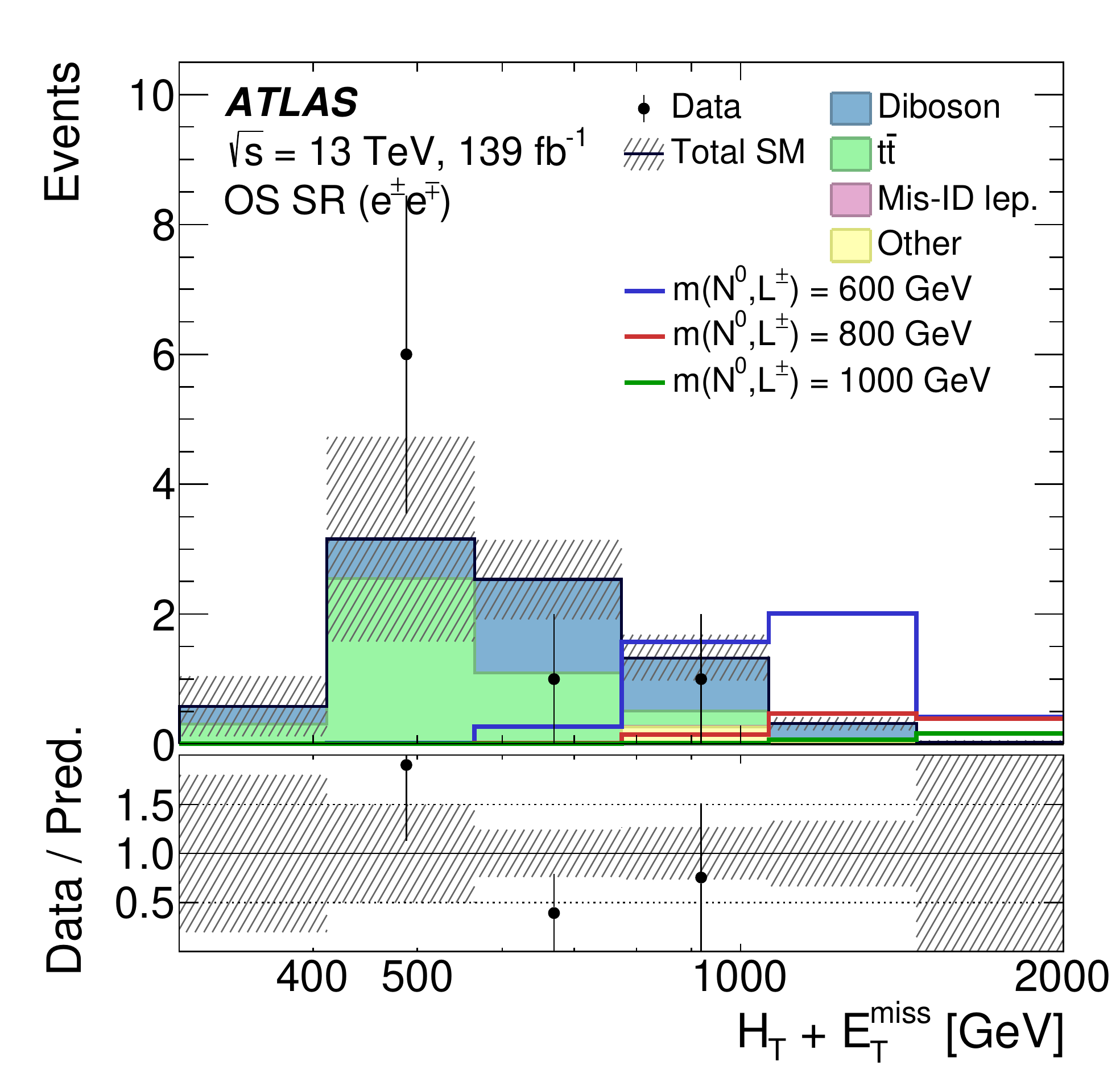}
\label{fig:Fit_SignalOS-1}
}
\subfloat[]{
\includegraphics[width=0.48\textwidth]{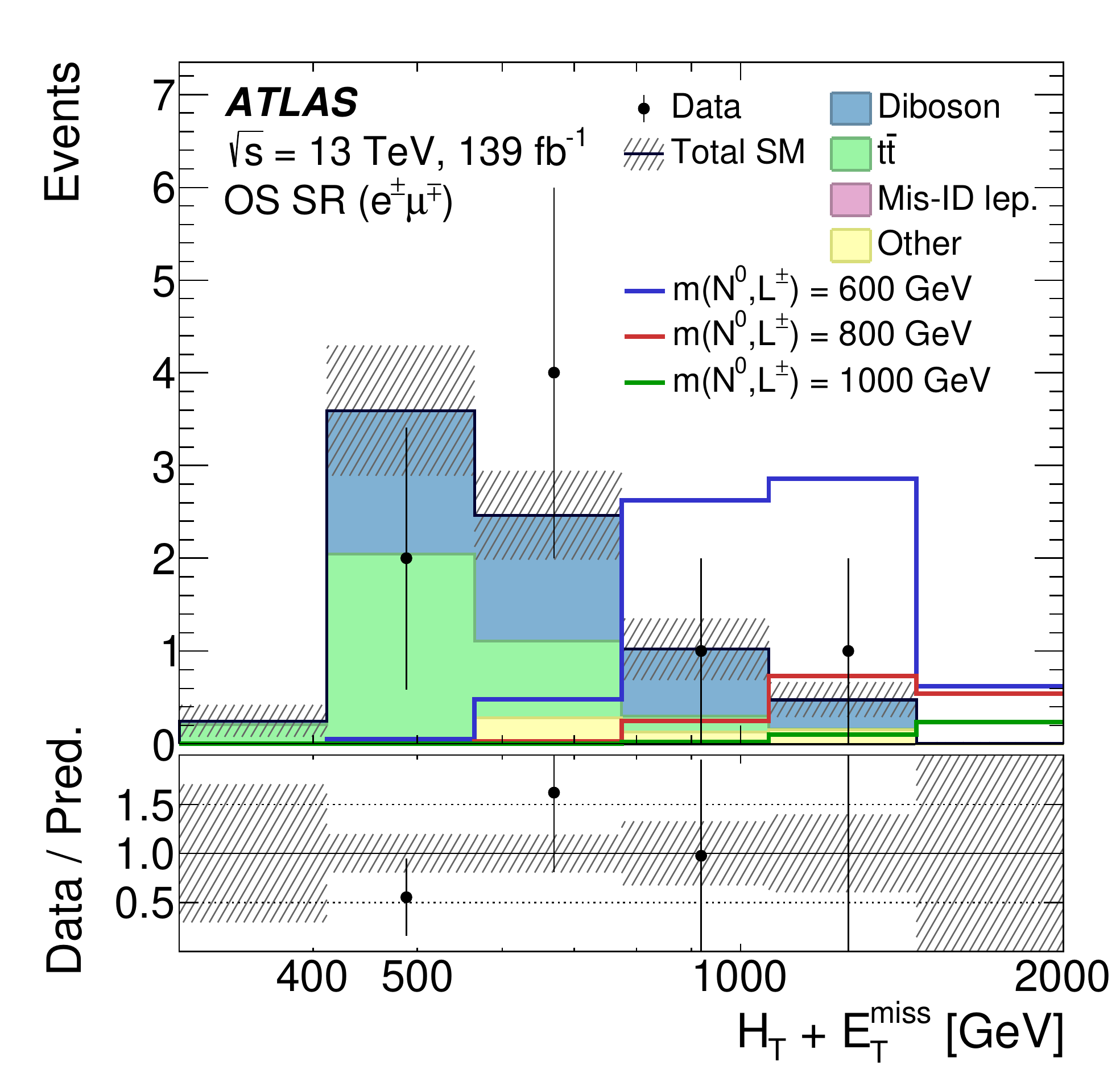}
\label{fig:Fit_SignalOS-2}
}
\\
\subfloat[]{
\includegraphics[width=0.48\textwidth]{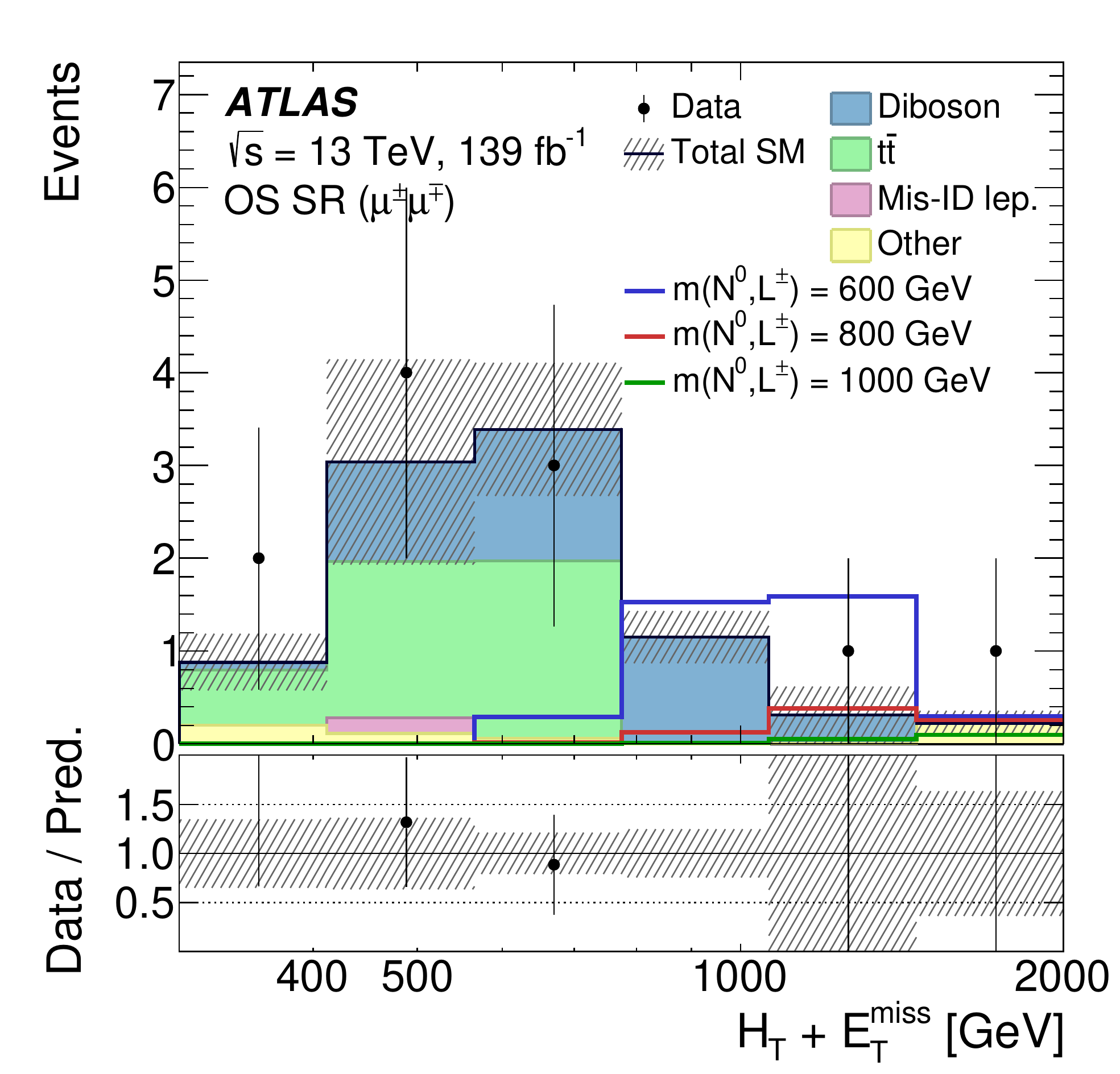}
\label{fig:Fit_SignalOS-3}
}
 
\caption{Distributions of \( \HT + \met \) in opposite-sign signal regions, namely
\protect\subref{fig:Fit_SignalOS-1}
the electron--electron signal region,  
\protect\subref{fig:Fit_SignalOS-2}
the electron--muon signal region, and  
\protect\subref{fig:Fit_SignalOS-3}
the muon--muon signal region after the background-only fit described in the text.  
The coloured lines correspond to signal samples with the \Nz and \Lpm
mass stated in the legend.
The hatched bands include all systematic uncertainties post-fit with
the correlations between various sources taken into account.
Errors on data are statistical only.
The lower panel shows the ratio of the observed data to the estimated
SM background.
}
\label{fig:Fit_SignalOS}
\end{figure}

\begin{figure}[htpb]
\centering
\subfloat[]{
\includegraphics[width=0.48\textwidth]{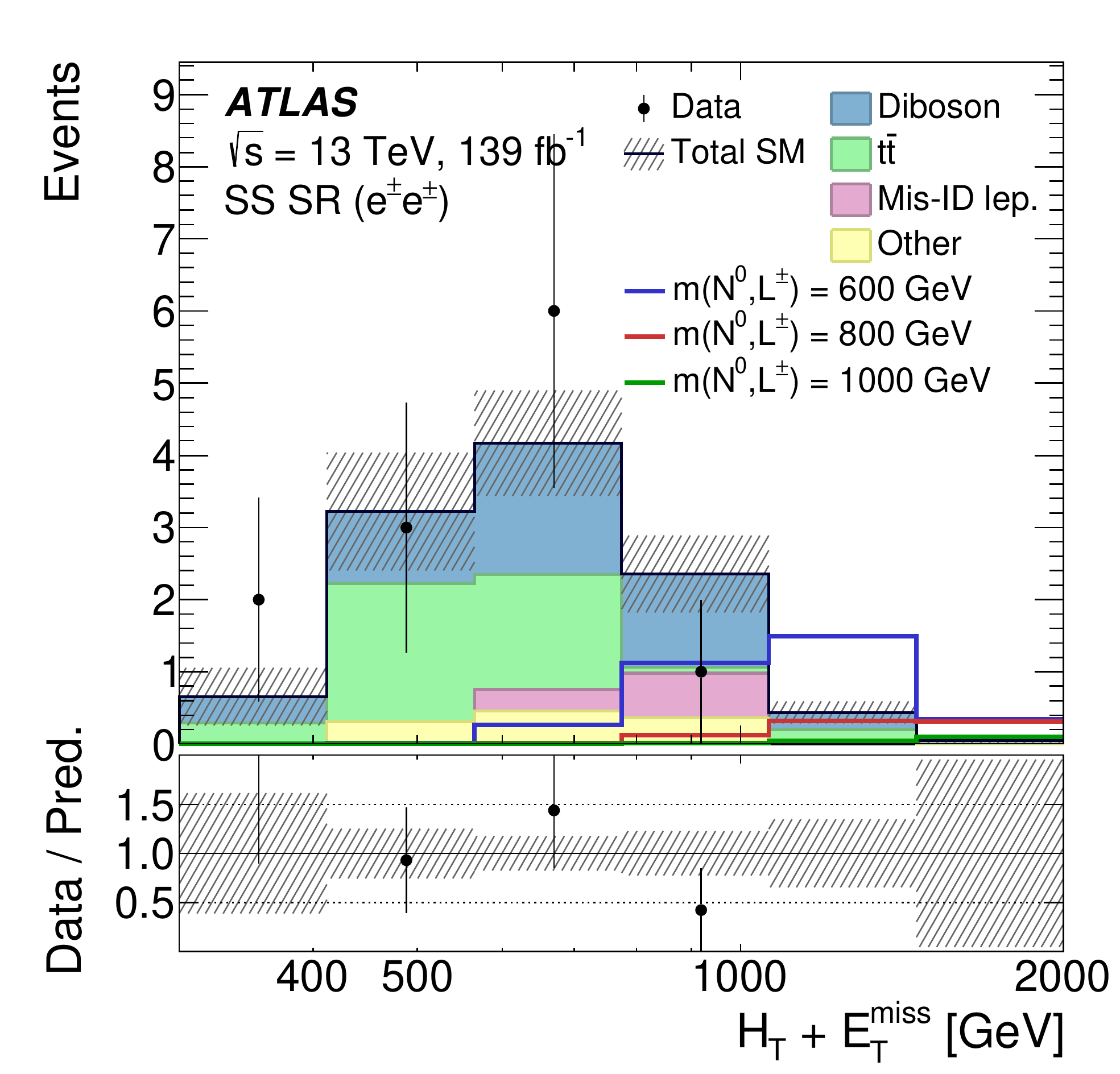}
\label{fig:Fit_SignalSS-1}
}
\subfloat[]{
\includegraphics[width=0.48\textwidth]{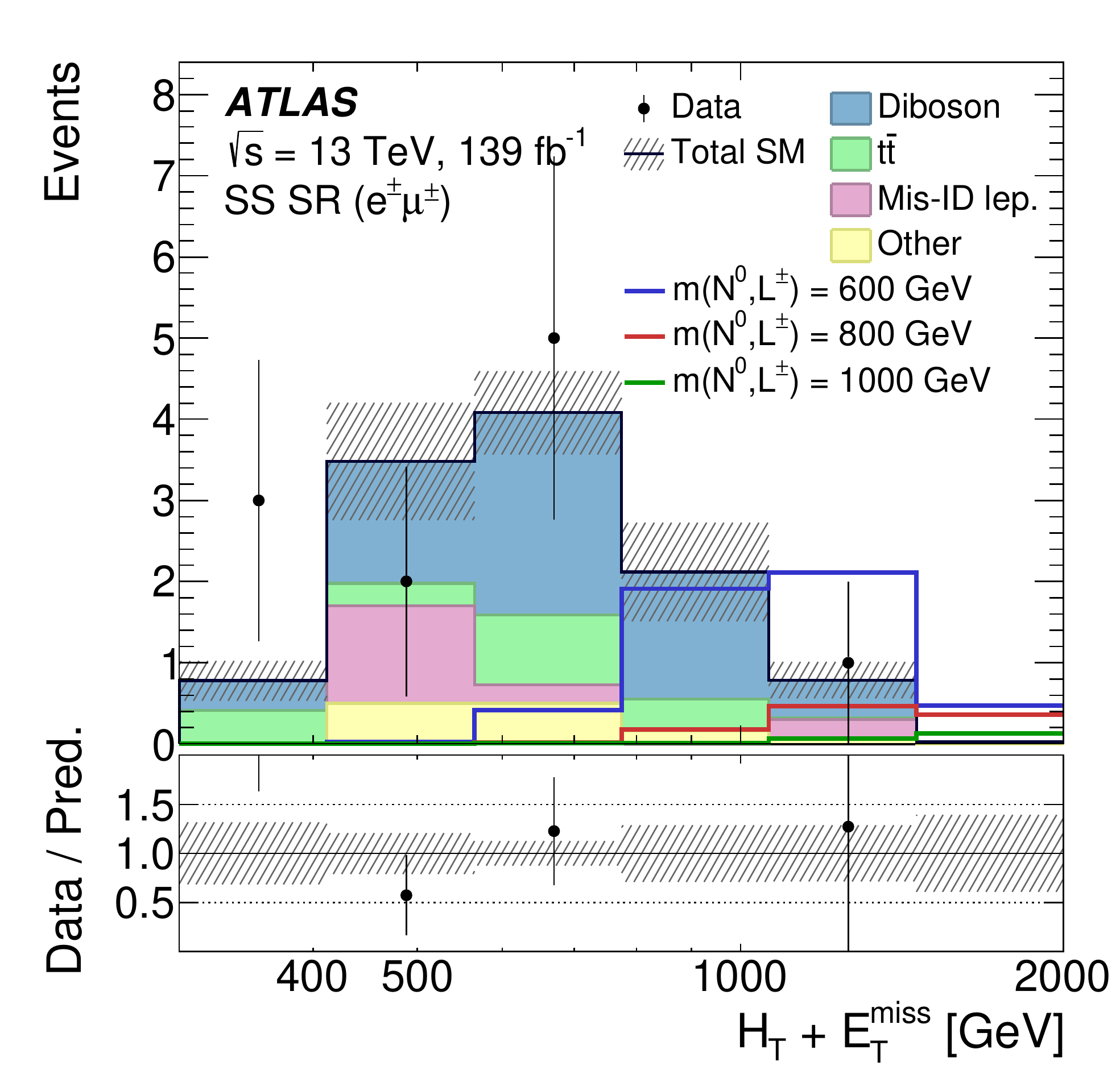}
\label{fig:Fit_SignalSS-2}
}
\\
\subfloat[]{
\includegraphics[width=0.48\textwidth]{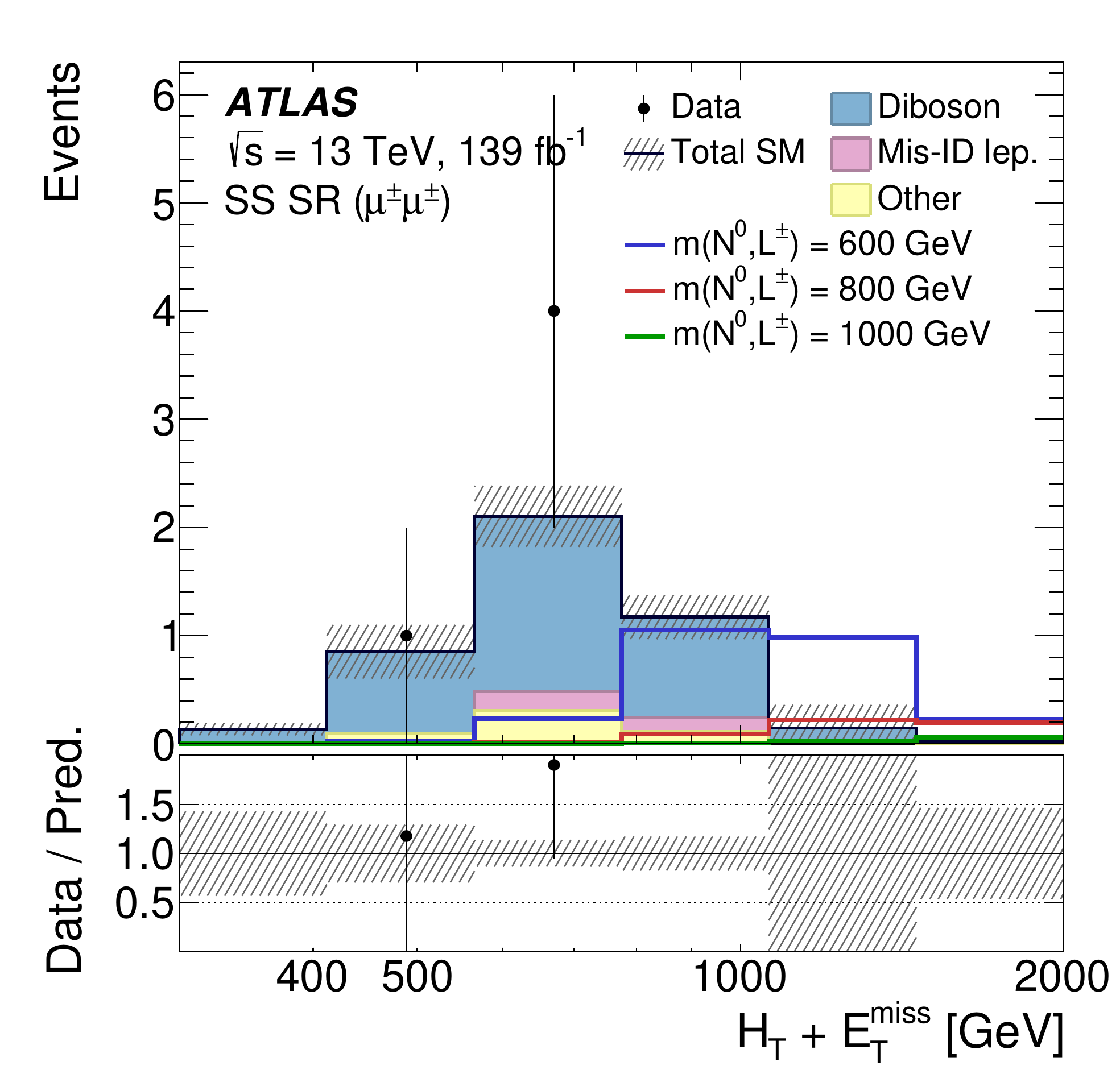}
\label{fig:Fit_SignalSS-3}
}
 
\caption{Distributions of \( \HT + \met \) in same-sign signal regions, namely
\protect\subref{fig:Fit_SignalSS-1}
the electron--electron signal region,  
\protect\subref{fig:Fit_SignalSS-2}
the electron--muon signal region, and  
\protect\subref{fig:Fit_SignalSS-3}
the muon--muon signal region after the background-only fit described in the text.  
The coloured lines correspond to signal samples with the \Nz and \Lpm
mass stated in the legend.
The hatched bands include all systematic uncertainties post-fit with
the correlations between various sources taken into account.
Errors on data are statistical only.
The lower panel shows the ratio of the observed data to the estimated
SM background.
}
\label{fig:Fit_SignalSS}
\end{figure}

Post-fit binned distributions of \( \HT + \met \) are shown in
\Figs{\ref{fig:Fit_SignalOS}}{\ref{fig:Fit_SignalSS}} for the signal regions. After the fit,
the compatibility between the data and the expected background is assessed
and good agreement is observed, with a  \(p\)-value of 0.5 for the background-only
hypothesis.\footnote{The \(p\)-value is defined as the probability that the observed difference between the data and the background prediction is a background fluctuation.}
In the absence of a significant deviation from expectations, \SI{95}{\%} confidence level (CL) upper limits on the  signal production cross-section are derived using the \( \mathrm{CL}_\mathrm{s} \)
method~\cite{Read:2002hq}. Pseudo-experiments are used to set the limit. All the plots and tables presented in this paper
are then derived from a likelihood fit assuming no signal presence, by fitting only the
background normalisation and corresponding sources of systematic uncertainty, as described in the text
(called a \textit{background-only fit}).

\begin{figure}[htpb]
\centering
\includegraphics[width=\textwidth]{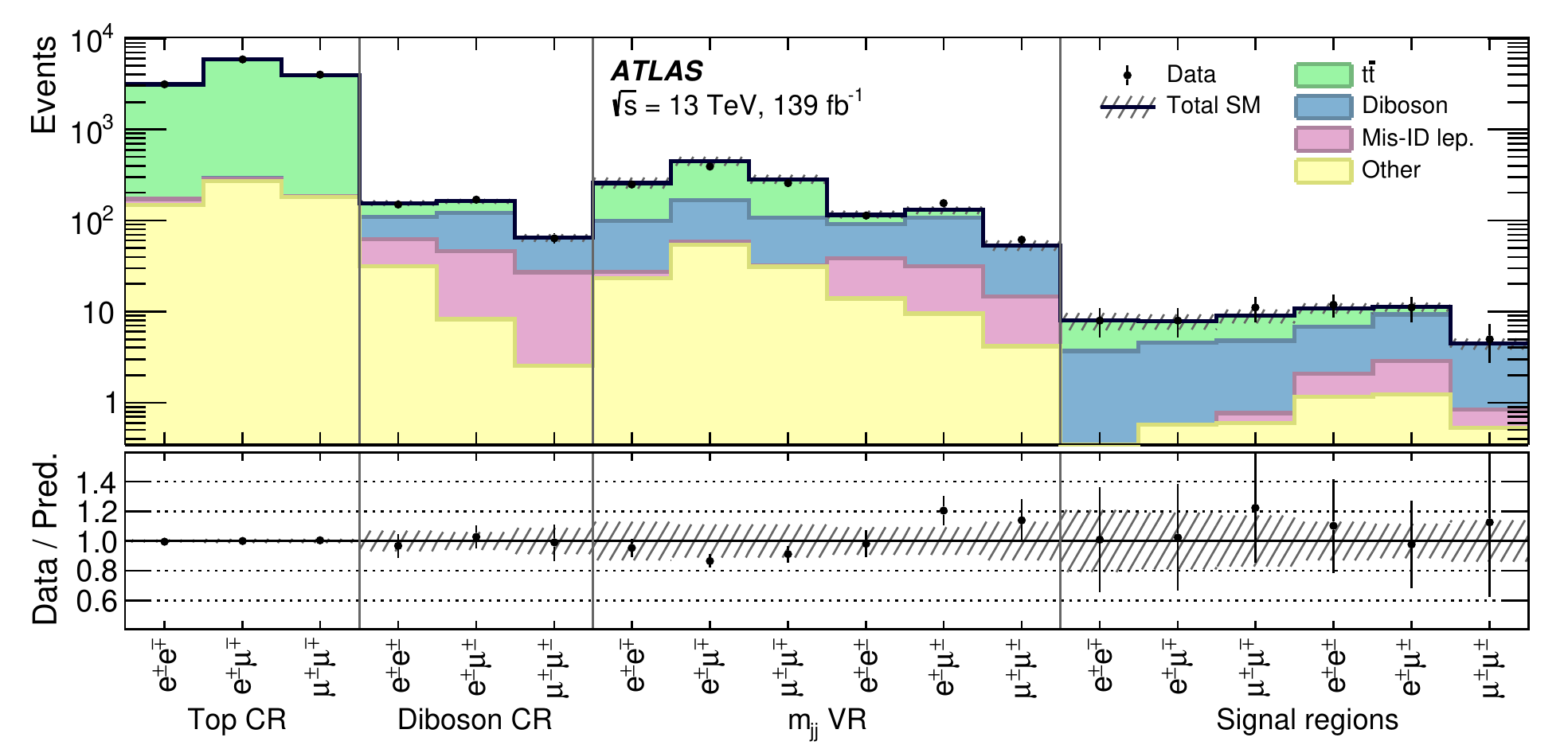}
\caption{The numbers of observed and expected events in the control, validation,
and signal regions for all channels, split by lepton flavour and
electric charge combination. The background expectation is the result
of the background-only fit described in the text. The hatched bands
include all post-fit systematic uncertainties with the correlations between
various sources taken into account. Errors on data are statistical only.
The lower panel shows the ratio of the observed data to the estimated
SM background.}
\label{fig:allRegions}
\end{figure}

\Fig{\ref{fig:allRegions}} shows good agreement, within uncertainties, between the expected background and observed events in all the regions and channels
considered in the analysis. The level of agreement between the data and background predictions
in the CRs and VRs, being well within the statistical and systematic uncertainties, demonstrates
the validity of the background estimation procedures.
 
\begin{figure}[htpb]
\centering
\includegraphics[width=\textwidth]{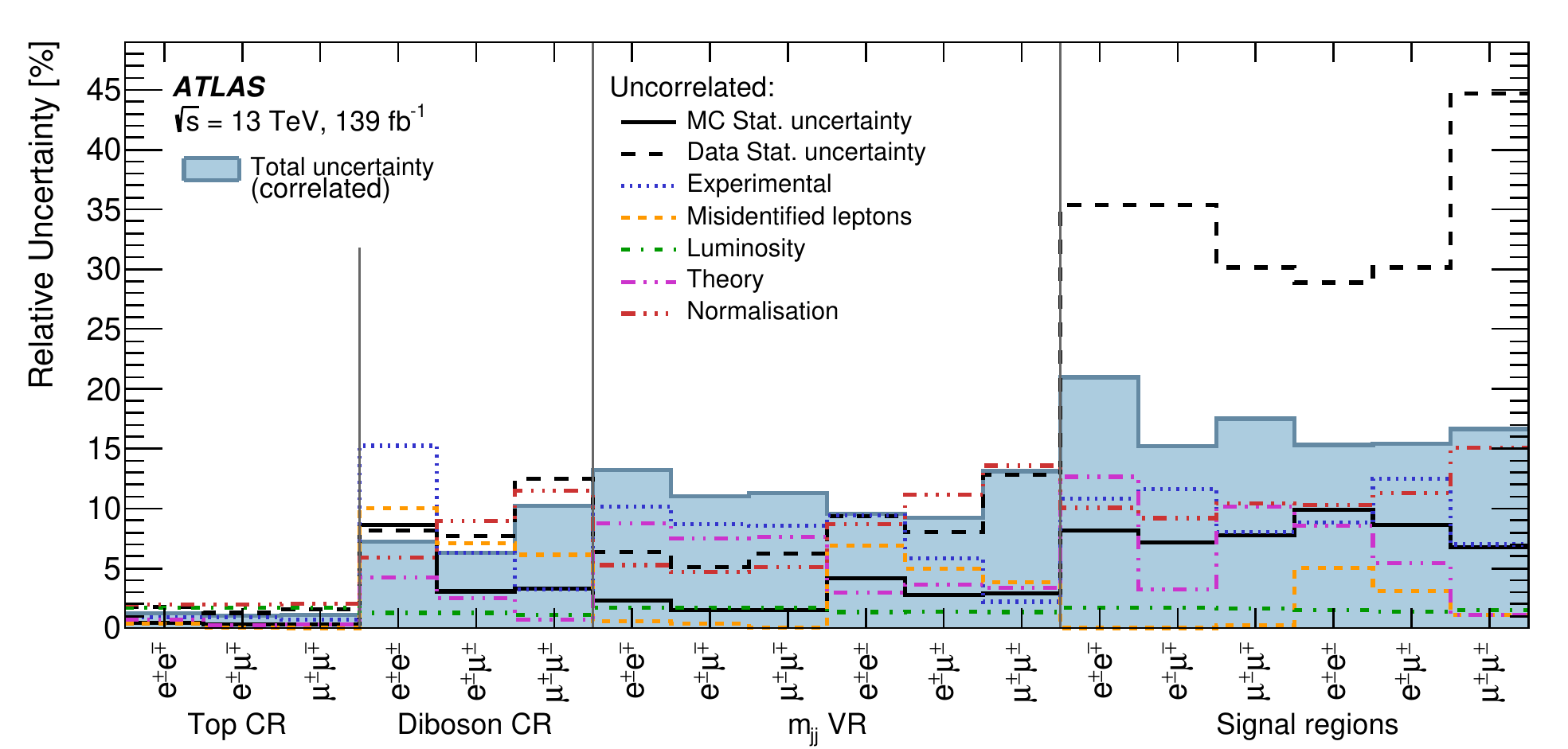}
\caption{Relative contributions of different sources of statistical
and systematic uncertainty in the total background yield estimation
after the fit.
Systematic uncertainties are calculated in an uncorrelated way by shifting in turn
only one nuisance parameter from the post-fit value by one standard deviation,
keeping all the other parameters at their post-fit values, and comparing
the resulting event yield with the nominal yield.
Validation regions are not included in the fit.
Individual uncertainties can be correlated, and do not necessarily add
in quadrature to the total background uncertainty, which is indicated by
`Total uncertainty'.}
\label{fig:allRegionsSys}
\end{figure}

The total relative systematic uncertainty in the background yields after the fit,
and its breakdown into components, is presented in \Fig{\ref{fig:allRegionsSys}}.
The contributions of theoretical and experimental uncertainties in the SR are both of the order of \SI{10}{\%}. The
dominant uncertainty is due to the limited number of simulated events.
It is reduced in the final fit combination, yielding a  total uncertainty of less than  \SI{20}{\%}.
After the fit, none of the nuisance parameters are pulled from their central values or have the uncertainties constrained.
 
Predicted numbers of background events in signal regions are listed together with the
observed number of events in data in \Tab{\ref{tab:Yields_Signal_2L}}.
Expected signal yields for several mass points are given for comparison.
 
\begin{table}[bp]
\caption{The number of expected
background events in signal regions after the likelihood fit,
compared with the data.
Uncertainties correspond to the total uncertainties in the predicted
event yields and can be smaller than the value of individual
contributions summed in quadrature because the latter can be anti-correlated.
Due to rounding, the totals can differ from the sums of components.
The \ttbar and diboson normalisations are allowed to float in the fit.
As a reference, representative signal yields and
their errors for signal masses \(m(\Nz, \Lpm)\), denoted by \(m\) in
the table, are shown in the bottom part of the table.}
\label{tab:Yields_Signal_2L}
 
\begin{center}
{
\small
\setlength{\tabcolsep}{3.5pt}
\begin{tabular}{l
S[table-format=1.4] @{\,}@{\( \pm \)}@{\,} S[table-format=1.4]
S[table-format=1.4] @{\,}@{\( \pm \)}@{\,} S[table-format=1.4]
S[table-format=1.2] @{\,}@{\( \pm \)}@{\,} S[table-format=1.2]
S[table-format=2.2] @{\,}@{\( \pm \)}@{\,} S[table-format=1.2]
S[table-format=2.2] @{\,}@{\( \pm \)}@{\,} S[table-format=1.2]
S[table-format=1.2] @{\,}@{\( \pm \)}@{\,} S[table-format=1.2] }
\toprule
Signal regions                               &
\multicolumn{2}{c}{OS \( e^\pm e^\mp \)}     &
\multicolumn{2}{c}{OS \( e^\pm \mu^\mp \)}   &
\multicolumn{2}{c}{OS \( \mu^\pm \mu^\mp \)} &
\multicolumn{2}{c}{SS \( e^\pm e^\pm \)}     &
\multicolumn{2}{c}{SS \( e^\pm \mu^\pm \)}   &
\multicolumn{2}{c}{SS \( \mu^\pm \mu^\pm \)}                                                                                                                                                                                                                 \\
\midrule
Observed events                              & \multicolumn{2}{l}{8} & \multicolumn{2}{l}{8} & \multicolumn{2}{l}{11} & \multicolumn{2}{l}{12} & \multicolumn{2}{l}{11} & \multicolumn{2}{l}{5}                                                              \\
\midrule
Total background                             & 7.9                   & 1.7                   & 7.8                    & 1.5                    & 9.0                    & 1.5                   & 10.9 & 1.4  & 11.3 & 1.3  & 4.44                    & 0.63 \\
\midrule
\ttbar                                       & 4.2                   & 2.5                   & 3.3                    & 1.2                    & 4.2                    & 1.2                   & 4.0  & 1.2  & 1.98 & 0.57 & \multicolumn{2}{l}{---}        \\
Diboson                                      & 3.4                   & 1.5                   & 3.99                   & 0.60                   & 4.03                   & 0.60                  & 4.8  & 1.0  & 6.4  & 1.3  & 3.60                    & 0.52 \\
Misidentified leptons                        & 0.0049                & 0.0005                & 0.0051                 & 0.0005                 & 0.17                   & 0.03                  & 0.92 & 0.35 & 1.65 & 0.34 & 0.31                    & 0.04 \\
Other                                        & 0.31                  & 0.27                  & 0.56                   & 0.29                   & 0.60                   & 0.45                  & 1.17 & 0.12 & 1.24 & 0.16 & 0.53                    & 0.20 \\
\midrule
Signal expectation                           & \multicolumn{2}{l}{}  & \multicolumn{2}{l}{}  & \multicolumn{2}{l}{}   & \multicolumn{2}{l}{}   & \multicolumn{2}{l}{}   & \multicolumn{2}{l}{}                                                               \\
\( m = \SI{600}{\GeV} \)                     & 4.26                  & 0.31                  & 6.64                   & 0.44                   & 3.71                   & 0.23                  & 3.24 & 0.27 & 4.94 & 0.34 & 2.53                    & 0.14 \\  
\( m = \SI{800}{\GeV} \)                     & 1.02                  & 0.10                  & 1.56                   & 0.15                   & 0.78                   & 0.06                  & 0.76 & 0.08 & 1.03 & 0.11 & 0.53                    & 0.04 \\  
\( m = \SI{1000}{\GeV} \)                    & 0.24                  & 0.03                  & 0.37                   & 0.04                   & 0.17                   & 0.02                  & 0.15 & 0.02 & 0.22 & 0.03 & 0.10                    & 0.01 \\  
\bottomrule
\end{tabular}
}
\end{center}
\end{table}
 
\begin{figure}[tbp]
\centering
\includegraphics[width=0.7\textwidth]{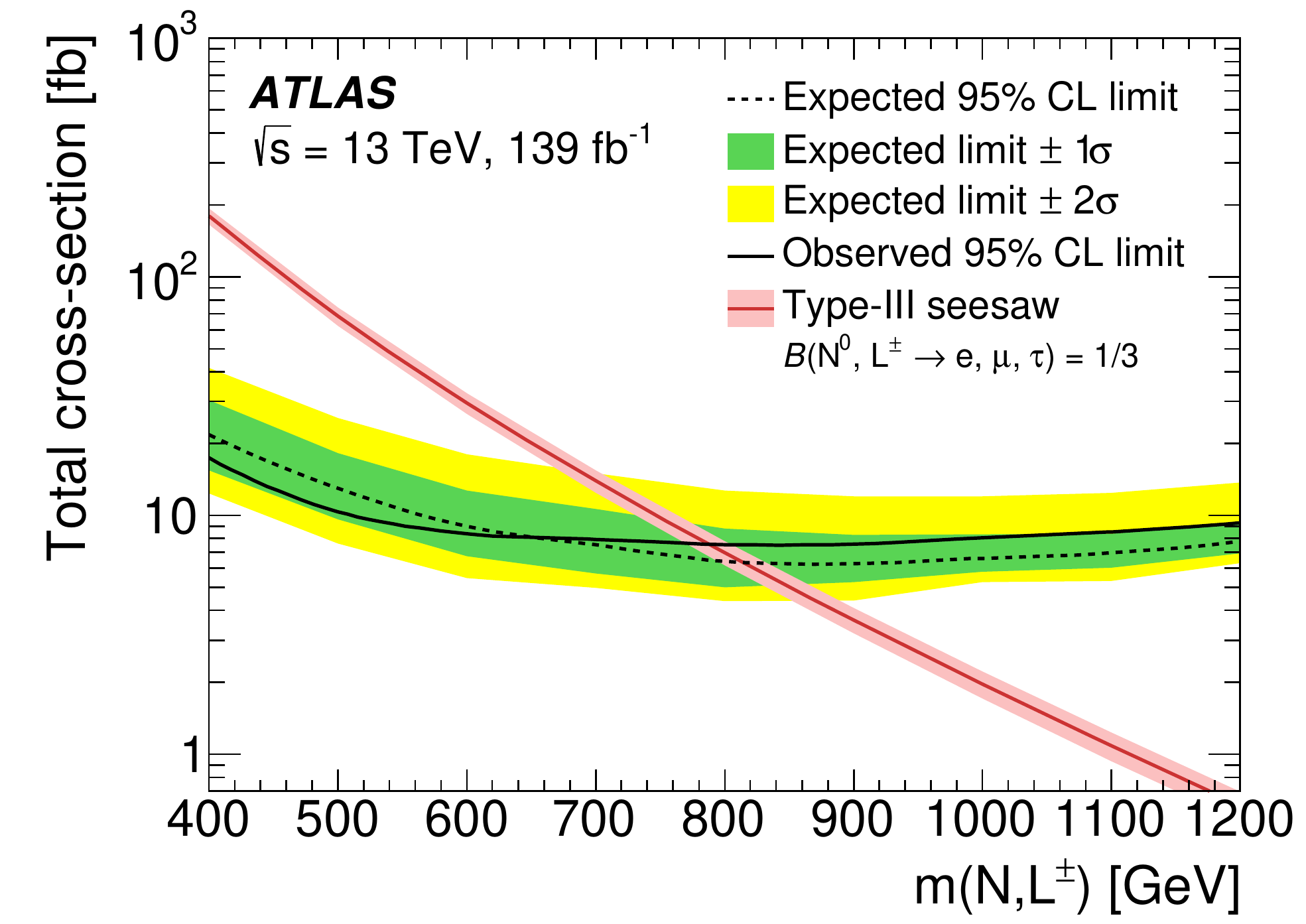}
\caption{Expected and observed \SI{95}{\%} \( \mathrm{CL}_\mathrm{s} \)
exclusion limits for the \typeIIIseesaw process with the corresponding
one- and two-standard-deviation bands, showing the \SI{95}{\%} CL upper limit
on the cross-section.
The theoretical signal cross-section prediction,
given by the NLO calculation~\cite{Fuks:2012qx,Fuks:2013vua}, is shown as a red line
with the corresponding uncertainty band.
}
\label{fig:exclusion}
\end{figure}

The  upper limits on the production cross-sections of the processes
\( pp \rightarrow W^{*} \rightarrow \Nz \Lpm \)
and
\( pp \rightarrow Z^{*} \rightarrow \Lpm \Lmp \)
at the \SI{95}{\%} CL are evaluated as a function of the heavy-lepton mass.
The resulting exclusion limits on the signal
cross-section are shown in \Fig{\ref{fig:exclusion}}.
By comparing the upper limits on the cross-section with the theoretical model dependence of the
cross-section on the heavy-lepton mass, the lower mass limit on the mass of the \typeIIIseesaw heavy leptons \Nz and \Lpm can be derived.
The expected lower limit is
\SI[parse-numbers=false]{820^{+40}_{-60}}{\GeV}, where the uncertainties
on the limit are extracted from the \( \pm\SI{1}{\sigma} \) band,
while the observed lower limit on the mass is \SI{790}{\GeV},
excluding the mass values below this point.
 
\FloatBarrier
 
\section{Conclusion}
\label{sec:conclusion}
 
The ATLAS detector at the Large Hadron Collider was used to search for
the pair production of heavy leptons predicted by the \typeIIIseesaw model.
The analysis is performed using a final state containing two same-charge or
opposite-charge leptons, electrons or muons, two jets from a hadronically
decaying \Wboson boson that were not identified as \(b\)-tagged, and large missing transverse momentum.
The search uses \SI{139}{\ifb} of data from proton–proton collisions at
\( \sqrt{s} = \SI{13}{\TeV} \), recorded during the 2015, 2016, 2017 and 2018
data-taking periods.
 
No significant excess above the SM prediction was found.
Limits are set on the \typeIIIseesaw heavy-lepton masses, using the simplified
\typeIIIseesaw model and assuming branching fractions to all lepton flavours
to be equal. Heavy leptons with masses below \SI{790}{\GeV} are excluded at the
\SI{95}{\%} confidence level with the expected lower mass limit at
\SI[parse-numbers=false]{820^{+40}_{-60}}{\GeV}.

\section*{Acknowledgements}
 

We thank CERN for the very successful operation of the LHC, as well as the
support staff from our institutions without whom ATLAS could not be
operated efficiently.
 
We acknowledge the support of ANPCyT, Argentina; YerPhI, Armenia; ARC, Australia; BMWFW and FWF, Austria; ANAS, Azerbaijan; SSTC, Belarus; CNPq and FAPESP, Brazil; NSERC, NRC and CFI, Canada; CERN; ANID, Chile; CAS, MOST and NSFC, China; COLCIENCIAS, Colombia; MSMT CR, MPO CR and VSC CR, Czech Republic; DNRF and DNSRC, Denmark; IN2P3-CNRS and CEA-DRF/IRFU, France; SRNSFG, Georgia; BMBF, HGF and MPG, Germany; GSRT, Greece; RGC and Hong Kong SAR, China; ISF and Benoziyo Center, Israel; INFN, Italy; MEXT and JSPS, Japan; CNRST, Morocco; NWO, Netherlands; RCN, Norway; MNiSW and NCN, Poland; FCT, Portugal; MNE/IFA, Romania; JINR; MES of Russia and NRC KI, Russian Federation; MESTD, Serbia; MSSR, Slovakia; ARRS and MIZ\v{S}, Slovenia; DST/NRF, South Africa; MICINN, Spain; SRC and Wallenberg Foundation, Sweden; SERI, SNSF and Cantons of Bern and Geneva, Switzerland; MOST, Taiwan; TAEK, Turkey; STFC, United Kingdom; DOE and NSF, United States of America. In addition, individual groups and members have received support from BCKDF, CANARIE, Compute Canada, CRC and IVADO, Canada; Beijing Municipal Science \& Technology Commission, China; COST, ERC, ERDF, Horizon 2020 and Marie Sk{\l}odowska-Curie Actions, European Union; Investissements d'Avenir Labex, Investissements d'Avenir Idex and ANR, France; DFG and AvH Foundation, Germany; Herakleitos, Thales and Aristeia programmes co-financed by EU-ESF and the Greek NSRF, Greece; BSF-NSF and GIF, Israel; La Caixa Banking Foundation, CERCA Programme Generalitat de Catalunya and PROMETEO and GenT Programmes Generalitat Valenciana, Spain; G\"{o}ran Gustafssons Stiftelse, Sweden; The Royal Society and Leverhulme Trust, United Kingdom.
 
The crucial computing support from all WLCG partners is acknowledged gratefully, in particular from CERN, the ATLAS Tier-1 facilities at TRIUMF (Canada), NDGF (Denmark, Norway, Sweden), CC-IN2P3 (France), KIT/GridKA (Germany), INFN-CNAF (Italy), NL-T1 (Netherlands), PIC (Spain), ASGC (Taiwan), RAL (UK) and BNL (USA), the Tier-2 facilities worldwide and large non-WLCG resource providers. Major contributors of computing resources are listed in Ref.~\cite{ATL-SOFT-PUB-2020-001}.
 

\printbibliography
 
\clearpage
 
\begin{flushleft}
\hypersetup{urlcolor=black}
{\Large The ATLAS Collaboration}

\bigskip

\AtlasOrcid[0000-0002-6665-4934]{G.~Aad}$^\textrm{\scriptsize 102}$,    
\AtlasOrcid[0000-0002-5888-2734]{B.~Abbott}$^\textrm{\scriptsize 128}$,    
\AtlasOrcid{D.C.~Abbott}$^\textrm{\scriptsize 103}$,    
\AtlasOrcid[0000-0002-2788-3822]{A.~Abed~Abud}$^\textrm{\scriptsize 36}$,    
\AtlasOrcid[0000-0002-1002-1652]{K.~Abeling}$^\textrm{\scriptsize 53}$,    
\AtlasOrcid[0000-0002-2987-4006]{D.K.~Abhayasinghe}$^\textrm{\scriptsize 94}$,    
\AtlasOrcid[0000-0002-8496-9294]{S.H.~Abidi}$^\textrm{\scriptsize 167}$,    
\AtlasOrcid[0000-0002-8279-9324]{O.S.~AbouZeid}$^\textrm{\scriptsize 40}$,    
\AtlasOrcid{N.L.~Abraham}$^\textrm{\scriptsize 156}$,    
\AtlasOrcid[0000-0001-5329-6640]{H.~Abramowicz}$^\textrm{\scriptsize 161}$
% ...
\bigskip
\\

% ...
$^{36}$CERN, Geneva; Switzerland.\\
% ...
$^{40}$Niels Bohr Institute, University of Copenhagen, Copenhagen; Denmark.\\
% ...
$^{53}$II. Physikalisches Institut, Georg-August-Universit\"{a}t G\"ottingen, G\"ottingen; Germany.\\
% ...
$^{94}$Department of Physics, Royal Holloway University of London, Egham; United Kingdom.\\
% ...
$^{102}$CPPM, Aix-Marseille Universit\'e, CNRS/IN2P3, Marseille; France.\\
$^{103}$Department of Physics, University of Massachusetts, Amherst MA; United States of America.\\
% ...
$^{128}$Homer L. Dodge Department of Physics and Astronomy, University of Oklahoma, Norman OK; United States of America.\\
% ...
$^{156}$Department of Physics and Astronomy, University of Sussex, Brighton; United Kingdom.\\
% ...
$^{161}$Raymond and Beverly Sackler School of Physics and Astronomy, Tel Aviv University, Tel Aviv; Israel.\\
% ...
$^{167}$Department of Physics, University of Toronto, Toronto ON; Canada.\\
% ...

\end{flushleft}


\end{document}